\documentclass[%
 reprint,
superscriptaddress,
 amsmath,amssymb,
 aps,
]{revtex4-2}

\usepackage{graphicx}
\usepackage{dcolumn}
\usepackage{bm}

\begin{document}

\preprint{APS/123-QED}

\title{Collective nature of high-$Q$ resonances in finite-size photonic metastructures}


\author{Thanh Xuan Hoang}
 \email{hoangtx@ihpc.a-star.edu.eg}
\affiliation{Institute of High Performance Computing (IHPC), Agency for Science, Technology and Research (A$^\star$STAR), 1 Fusionopolis Way, \#16-16 Connexis, Singapore 138632, Republic of Singapore}

\author{Daniel Leykam}%
\email{daniel.leykam@gmail.com}
\affiliation{Centre for Quantum Technologies, National University of Singapore, 3 Science Drive 2, Singapore 117543}%

\author{Hong-Son Chu}%
\author{Ching Eng Png}%
\affiliation{Institute of High Performance Computing (IHPC), Agency for Science, Technology and Research (A$^\star$STAR), 1 Fusionopolis Way, \#16-16 Connexis, Singapore 138632, Republic of Singapore}

\author{Francisco J. Garc\'{i}a-Vidal}%
\affiliation{Institute of High Performance Computing (IHPC), Agency for Science, Technology and Research (A$^\star$STAR), 1 Fusionopolis Way, \#16-16 Connexis, Singapore 138632, Republic of Singapore}
\affiliation{Departamento de F\'{i}sica Te\'{o}rica de la Materia Condensada and Condensed Matter Physics Center (IFIMAC), Universidad Aut\'{o}noma de Madrid, E-28049 Madrid, Spain}%

\author{Yuri S. Kivshar}%
\email{yuri.kivshar@anu.edu.au}
\affiliation{Nonlinear Physics Centre, Research School of Physics, The Australian National University, Canberra ACT 2601, Australia}%

\date{\today}

\begin{abstract}
We study high quality-factor (high $Q$) resonances supported by periodic arrays of Mie resonators from the perspectives of both Bloch wave theory and multiple scattering theory. We reveal that, unlike a common belief,  the bound states in the continuum (BICs) derived by the Bloch-wave theory do not directly determine the resonance with the highest $Q$ value in large but finite arrays. 
Higher $Q$ factors appear to be associated with collective resonances formed by nominally guided modes below the light line associated with strong effect of both electric and magnetic multipoles. Our findings offer valuable insights into accessing the modes with higher $Q$ resonances via bonding modes within finite metastructures. Our results underpin the pivotal significance of magnetic and electric multipoles in the design of resonant metadevices and nonlocal flat-band optics. Moreover, our demonstrations reveal that coupled arrays of high-$Q$ microcavities do not inherently result in a stronger light-matter interaction when compared to coupled low-$Q$ nanoresonators. This result emphasizes the critical importance of the study of multiple light-scattering effects in cavity-based systems.
\end{abstract}

\maketitle


\section{\label{Introduction}Introduction}

High quality factor $Q$ nanophotonic resonances are important for various applications ranging from photonic crystal cavities for quantum photonics~\cite{Mahmoodian:17,Lu2022,gonzalez2024light} to metasurfaces for ultra-thin optical beam-shaping elements~\cite{kivshar2022rise,shastri2023nonlocal,schiattarella2024directive}. The former are based on nominally-infinite $Q$ guided modes of photonic crystal slabs, while the latter employ finite $Q$ Mie resonances of wavelength-scale particles. Remarkably, fine-tuning or special symmetries have been predicted to convert low $Q$ resonances into infinite $Q$ modes known as bound states in the continuum (BICs)~\cite{hsu2016bound,kang2023applications}. While there has been enormous interest in the BIC concept, theoretical predictions regarding BICs as cavities with infinite $Q$ diverge from practical implementations, with experimentally-measured $Q$ values limited to less than one million~\cite{jin2019topologically}.

Both the photonic crystal cavity and BIC approaches are inspired by analogies between matter and light waves. These concepts are rooted in the physics of scattering-free propagation observed in Bloch waves within infinite periodic photonic crystals~\cite{Joan2008,hsu2016bound}. Discrepancies between the nominal and measured $Q$ values are typically attributed to enhanced scattering losses arising from fabrication imperfections or finite sample sizes~\cite{jin2019topologically,hwang2021ultralow,chen2022observation}, which are both neglected in the Bloch wave theory~\cite{hsu2016bound}. These effects become more important as devices are scaled down, leading to much lower $Q$ factors reported in nanophotonic systems compared to single microcavities, which support whispering-gallery modes with measurable $Q$ in the billions~\cite{Wu:20}. Thus, while the BIC approach gives an elegant and intuitive way to understand the $Q$ factors of Bloch waves, it lacks quantitative predictive power for real finite size systems.

\begin{figure*}[htbp]
\includegraphics[width = 16 cm]{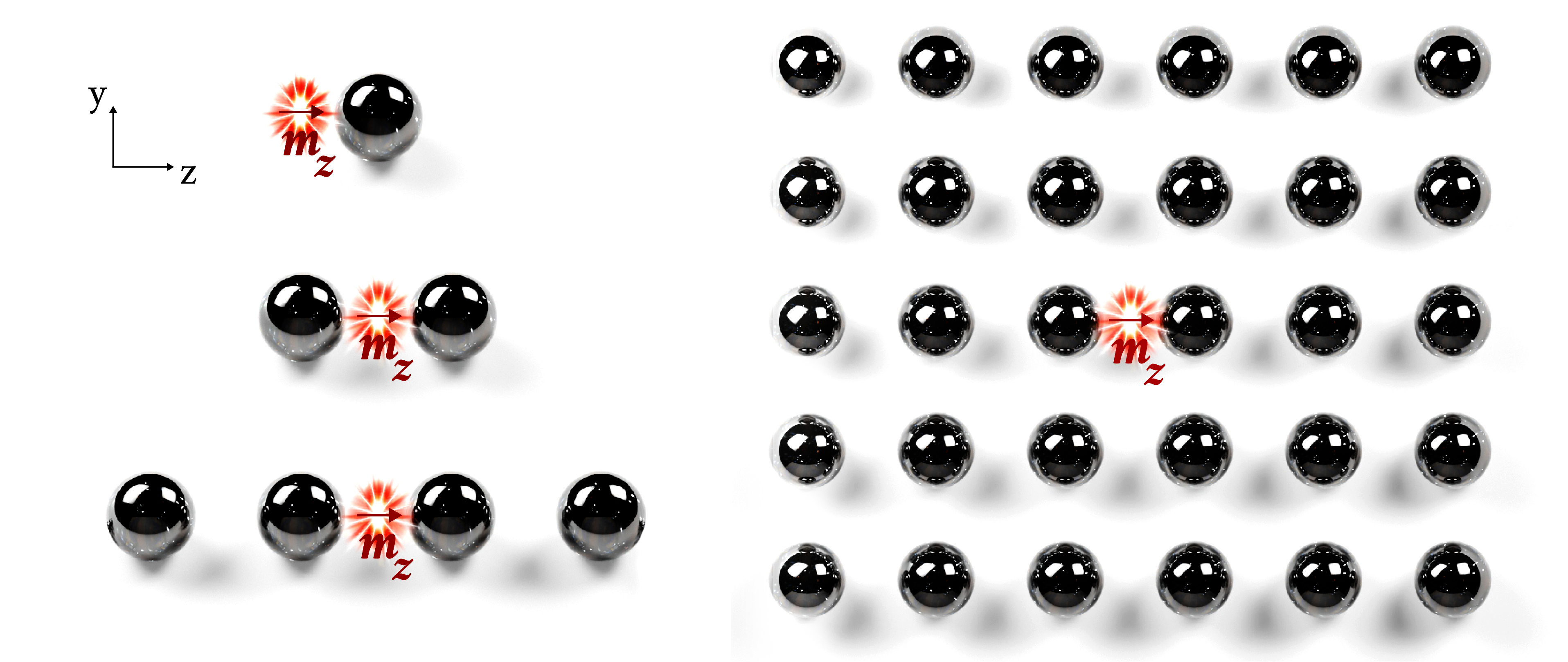}
\caption{\label{F1} Schematic illustrating the interaction of a magnetic dipole with various structures: a meta-atom, a diatomic meta-molecule, a linear chain of spheres, and a two-dimensional (2D) array of spheres. Extending the linear chain and the 2D array to infinite periodic structures transitions them into photonic crystals capable of supporting Bloch waves, including bound states within the continuum.}
\end{figure*}

Here we study resonances supported by arrays of Mie-resonant nanoparticles from the viewpoint of multiple scattering theory (MST), schematically illustrated in Fig.~\ref{F1}. We show that the scattering wave viewpoint provides a simple way to understand the emergence of (quasi-)BICs and other high $Q$ resonances  in metastructures and metasurfaces in terms of collective resonances whose $Q$ scales with the system size, diverging in the limit of an infinite system. Intriguingly, our findings reveal that the collective resonances of coupled high-$Q$ microcavities do not necessarily result in Q factor divergence. This is in contrast to the pronounced divergence observed when coupling low-$Q$ Mie resonators, highlighting the intricate physics involved in strong multiple scattering effects.

In the following we will consider Mie-resonant silicon nanoparticles with a fixed sphere radius of 210 nm and a refractive index of 3.5. These parameters ensure that the high $Q$ collective resonances fall within the crucial near-infrared frequency range, essential for various technological applications~\cite{hoang2022high,couteau2023applications}, and are similar to recent experiments studying quasi-BICs~\cite{kodigala2017lasing,ha2018directional,liu2019high}.

\section{Bloch waves vs. scattering waves}

In the weak coupling regime, the emission rate $\gamma$ of a light emitter interacting with the local three-dimensional (3D) electromagnetic environment is described by the multipolar interaction Hamiltonian:
\begin{equation}
H_{\text{int}} = - \bm{\mu}\cdot{\bf E}({\bf r}_{\mu},t) - \bm{m}\cdot{\bf B}({\bf r}_{m},t)-\dots \label{E1}
\end{equation}
Here, $\bm{m}$ and ${\bf r}_{m}$ ($\bm{\mu}$ and ${\bf r}_{\mu}$) represent the magnetic (electric) dipole moment and position of the dipole, respectively. Equation~\eqref{E1} is applicable to investigate both the dynamical and stationary properties of the emission rate. Our focus lies on the stationary modes of BICs, where we employ predefined magnetic dipoles with constant moments across the entire frequency range of interest.

For systems involving magnetic dipole transitions at equilibrium, the emission rate is
\begin{equation}
\gamma =\frac{\omega}{2}\text{Im}\{\bm{m}^\ast\cdot{\bf B}({\bf r}_{m})\}. \label{E2}
\end{equation}
Here, $\omega$ represents the angular frequency associated with the dipolar transition. As our interest lies in the enhancement of $\gamma$ due to emitter-environment interactions, we utilize the Purcell factor defined as:
\begin{equation}
F_P =\frac{\gamma}{\gamma_0}, \label{E3}
\end{equation}
where $\gamma_0$ denotes the emission rate of the emitter in the corresponding homogeneous material, which is vacuum in Fig. \ref{F1}.

Equations \eqref{E2} and \eqref{E3} can effectively describe the interaction of a dipole with both Bloch waves and scattering waves, which correspond to different field profiles ${\bf B}({\bf r}_{m})$. The unique boundary conditions applied in solving Maxwell's equations for ${\bf B}({\bf r}_{m})$ result in significant distinctions between the two.

\subsection{Bloch waves: Diffracted waves as open channels} 

In the case of Bloch waves, the electromagnetic field satisfies
\begin{equation}
\mathbf{B}(\mathbf{r}+\mathbf{R}_{nu})=e^{i\mathbf{k}_B\cdot\mathbf{R}_{nu}}\mathbf{B}(\mathbf{r}),\label{E4}
\end{equation}
where $\mathbf{R}_{nu}$ is the vector connecting the $n$th and $u$th unit cells. Numerical simulations are employed to determine the magnetic field $\mathbf{B}(\mathbf{r})$ within the unit cell, utilizing an excitation source at the temporal frequency $f$. For the linear chain in Fig. \ref{F1}, Bloch boundary conditions are applied in the $z$ direction, while perfectly matched layers simulate outgoing waves in the $x$ and $y$ directions. These outgoing waves propagate in vacuum with a wavenumber $k=2\pi f/c$, $c$ being the speed of light. The dispersion relation $k_B(f)$ is computed based on these simulations, analyzing the magnetic field eigenmode $\mathbf{B}(\mathbf{r})$.

For a given $f$, if $k_B>k$, the Bloch wave is termed a guided mode~\cite{johnson1999guided}, aligning with a band below the light line in the band diagram. Conversely, if $k_B<k$, the Bloch wave couples with a finite number of diffracted channels, also known as open channels. These open channels collectively form what is referred to as a continuum due to their continuous spectra. The interaction with these open channels causes the bands above the light line to exhibit behavior akin to leaky resonances, leading to the concept of guided resonances~\cite{fan2002analysis}.

Subsequent advancements introduced the term `bound states' for guided bands, suggesting that Bloch waves -- propagating without scattering -- are truly `bound.' However, the finite nature of one-dimensional (1D) or two-dimensional (2D) photonic crystals inevitably leads to a complex interplay between Bloch waves and scattering waves in both theoretical and experimental realms~\cite{vaishnav2007matter}. These complexities are further compounded by recent works in guided resonances, giving rise to Bloch BICs with a definition differing from the original matter BICs~\cite{hsu2013observation,zhen2014topological,bulgakov2017topological}.

Despite its inherent existence, the scattering aspects in the interaction between light and 1D and 2D photonic crystals have largely been neglected. Scattering effects account for numerous fundamental phenomena, wherein the often-overlooked differences between light waves and matter waves play crucial roles~\cite{lagendijk1996resonant}. 

\subsection{Scattering waves: Partial multipole and plane waves as open channels}

The scattering field can be expanded using multipole and plane-wave representations  as~\cite{devaney1974multipole,hoang2014multipole}
\begin{widetext}
\begin{align}
&{\bf E}({\bf r})= 
 \sum_{l=1}^{L_{\text max}}\sum_{m=-l}^{l}\left[\alpha_{lm}{\bf N}_{lm}(k{\bf r})+\beta_{lm}{\bf M}_{lm}(k{\bf r})\right]=\frac{ik}{2\pi}\int_0^{2\pi}\,d\beta\int_{C^{\pm}}\,d\alpha\sin\alpha\hat{E}(\hat s)e^{i{\bm k}\cdot{\bm r}}, \label{E5}\\
 &{\bf H}({\bf r})= 
 \sum_{l=1}^{L_{\text max}}\sum_{m=-l}^{l}\left[\beta_{lm}{\bf N}_{lm}(k{\bf r})-\alpha_{lm}{\bf M}_{lm}(k{\bf r})\right]=\frac{ik}{2\pi}\int_0^{2\pi}\,d\beta\int_{C^{\pm}}\,d\alpha\sin\alpha\hat{H}(\hat s)e^{i{\bm k}\cdot{\bm r}}. \label{E6}
\end{align}
\end{widetext}
Here, ${\bf N}_{lm}$ and ${\bf M}_{lm}$ represent electric and magnetic multipole fields behaving as outgoing waves at infinity, while $\alpha_{lm}$ and $\beta_{lm}$ denote multipole expansion coefficients (MECs) specific to the interaction configuration. The spectral amplitude vectors $\hat{E}(\hat{s})$ and $\hat{H}(\hat{s})$ relate to the MECs, where $\hat{s}(\alpha,\beta) = \mathbf{k}/k$ is a unit vector with the polar angle $\alpha$ containing both real and complex values along the integration contours $C^{\pm}$. For convenience, we will use ${\bf H}={\bf B}/\mu_0$ to refer to the magnetic field. 

The complex values of $\alpha$ correspond to evanescent plane waves that decay exponentially in specific directions. Traditionally termed closed channels, these evanescent waves stand in contrast to their propagating counterparts, corresponding to the real values of $\alpha$, known as open channels~\cite{kodigala2017lasing}. In the context of the plane-wave expansion, emitters depicted in Fig. \ref{F1} interact with both propagating and evanescent wave components. Despite far-field analysis typically overlooking evanescent waves, these emitters interact with both types, necessitating energy dissipation into the far-field region to maintain system equilibrium.

The stored energy within evanescent waves doesn't represent a true BIC; upon deactivating the excitation source, the associated polarization energy gradually leaks into the far-field region. Notably, individual plane waves, extending infinitely and implying infinite energy, aren't physically realizable. Conversely, partial multipole fields offer a more comprehensible representation as physical waves. 

Due to the positive energy of light waves, each multipole mode always corresponds to a resonance, potentially possessing a substantial yet always finite quality factor (Q factor). These modes, operating within a 3D space, intricately engage with numerous degrees of freedom. Consequently, they produce far-field patterns characterized by varying intensities across different directions--displaying maximum and minimum intensities as dictated by their mode numbers. Crucially, all multipole modes remain open channels, perpetually interacting with their 3D environment. This absence of closed channels in their expansion aligns with the positive energy nature of light waves. These properties of multipole modes serve as fundamental components for dissecting the underlying physics governing phenomena such as symmetry-protected and accidental BICs.

The importance of the far-field can be seen through the calculation of the system's time-averaged radiated power, derived from the multipole expansion detailed in Eqs. \eqref{E5} and \eqref{E6}:
\begin{equation}
P = \frac{c}{8\pi}\sum_{l=1}^{L_{\text max}}\sum_{m=-l}^ll(l+1)\left[|\alpha_{lm}|^2+|\beta_{lm}|^2\right]. \label{E7}
\end{equation}
An alternative approach to Eq.~\eqref{E3} involves computing the Purcell factor as $F_P=P/P_0$, where $P_0$ denotes the radiating power in the corresponding homogeneous environment. Equation \eqref{E7} ensures the coexistence of both near and far fields. 

Further insight into the energy carried by individual plane waves can be captured by Eq. \eqref{E7}. For example, considering a plane wave traveling in the $z$ direction with circular polarization and an electric field ${\bf E}=(\hat{x}+i\hat{y})E_0e^{ikz}$, the associated MECs are
\begin{equation}
\alpha_{l;1}= i\beta_{l;1} = \frac{E_0}{k}i^{l+1}\sqrt{\frac{\pi(2l+1)}{l(l+1)}}.\label{E8}
\end{equation}
Using Eqs. \eqref{E7} and \eqref{E8}, the power carried by such a plane wave can be estimated as an infinite sum: $P=(cE_0^2)/(8k^2)\sum_{l=1}^{\infty}(2l+1)\rightarrow \infty$. This signifies the unphysical nature of a plane wave carrying infinite power in practical terms. Thus, multipole modes, acting as open channels, offer a more insightful framework for describing light phenomena in metastructures.

\section{Collective antibonding and bonding photonic modes }\label{S2}
 The theory of multiple Mie scattering allows us to express the field scattered by the \textit{$u$}-th sphere positioned at ${\bf r}_u$ in terms of a series of magnetic multipole fields ${\bf M}_{lm}$
\begin{equation}
{\bf E}_u({\bf r}_u) = \sum_{l=1}^{L_u}q_{l;0}^{(u)}{\bf M}_{l;0}(k[{\bf r}-{\bf r}_u]), \label{E9}
\end{equation}
where the required truncation order $L_u$ depends on the vacuum wavenumber $k$ and the spheres' radius; here
$L_u = 10$ is sufficient to obtain good agreement with a direct numerical solution of Maxwell’s equations (Lumerical FDTD). Due to the axial symmetry, the $m_z$ source excites only the magnetic multipole modes with $m=0$. The internal field of the \textit{$u$}-th sphere can be calculated in the MST in which we can relate its MECs $\eta^{(u)}_{l;0}$ to the MECs $q_{l;0}^{(u)}$ in Eq. \eqref{E9} via $\eta^{(u)}_{l;0} = q_{l;0}^{(u)}d_l^{(u)}/b_l^{(u)}$, where $b_l^{(u)}$ and $d_l^{(u)}$ are the Mie coefficients. Applying the MST results in the following equation taking into account all the short-range, long-range, and scattering couplings
\begin{equation}
q_{l^\prime;0}^{(n)}=b_{l^\prime}^{(n)}\left(A_{l^\prime;0}^{1;0}(\overrightarrow{OO_n})m_z+\sum_{u\neq n}\sum_{l=1}^{L_u}A_{l^\prime;0}^{l;0}(\overrightarrow{O_uO_n})q_{l;0}^{(u)}\right),\label{E10}
\end{equation}            
where $A_{l^\prime;0}^{l;0}(\overrightarrow{O_uO_n})$ translates the magnetic multipole field of ${\bf M}_{l;0}$ from the $u$-th sphere into the incident field approaching the $n$-th sphere \cite{hoang2017fano}. Equation \eqref{E10} allows us to compute all the MECs for studying the near- and far-field distributions as well as the Purcell factor. 

\subsection{Intrinsic magnetic quadrupole and octupole modes in photonic meta-atoms}
\begin{figure}[htbp]
\includegraphics[width = \columnwidth]{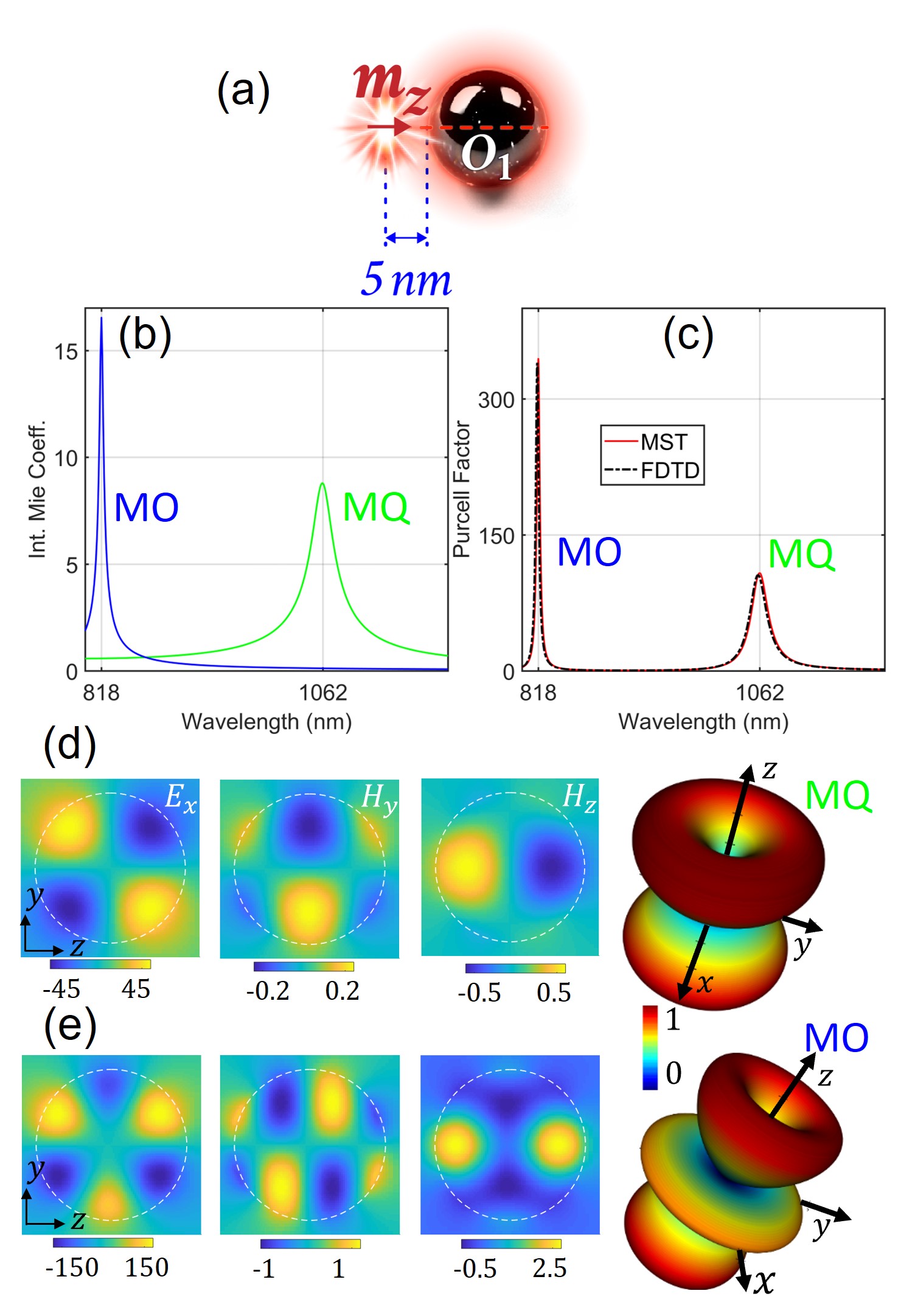}
\caption{\label{F2}(a) Interaction schematic of a magnetic dipole oriented along the $z$-axis interacting with a single sphere. (b) Spectral profiles of the internal Mie coefficients: $d_2$ (MQ) and $d_3$ (MO). (c) Purcell factor associated with the interaction schematic. The MST result based on Eq. \eqref{E7} closely matches the FDTD simulation using Eq.~\eqref{E3}. (d)-(e) Near- and far-field patterns correspond to the excitation of the magnetic quadrupole and octupole modes.}
\end{figure}
Figure \ref{F2} illustrates the spectral multipolar analyses detailing the interaction between a magnetic dipole and a single sphere. In Fig. \ref{F2}(b), the resonant wavelengths of the intrinsic magnetic quadrupole (MQ) and octupole (MO) are presented, with corresponding Q factors of 50 and 200, respectively. Generally, higher-order multipole modes exhibit higher Q factors, as observed here, owing to the dominance of whispering-gallery modes in the light confinement physics of a single sphere. These modes rely on the effect of partial internal reflection, which progressively approaches total internal reflection for higher-order multipole modes. In some cases, Q factors in the billions have been experimentally observed for sufficiently high orders. However, whispering-gallery cavities, despite their high Q factors, are often bulky and lack the nanoscale light manipulation capabilities characteristic of metastructures, which present distinct advantages in manipulating light properties at the nanoscale.

The spectral distribution of $F_P$ in Fig. \ref{F2}(c) reveals peak wavelengths consistent with those of the internal Mie coefficients, underscoring that Purcell enhancements are primarily driven by the intrinsic modes. To validate the accuracy of $F_P$ calculations, we conducted numerical simulations using a finite-difference time-domain (FDTD) commercial software (Lumerical FDTD).

The dominance of intrinsic modes in the interaction is further elucidated by the near- and far-field patterns shown in Figs. \ref{F2}(d) and \ref{F2}(e), corresponding to the MQ and MO modes, respectively. Notably, in the far-field distributions, the MQ mode exhibits a distinctive behavior characterized by destructive interference in the transverse $xy$ plane. This interference aligns with the anti-symmetric nature of the near-field $H_z$ component of the MQ mode, as depicted in Fig. \ref{F2}(d).

This observation concerning the interplay between near and far fields has often been foundational in arguing for the existence of symmetry-protected BICs. However, this argument encounters limitations when considering the symmetric nature of its electric $E_x$ component. Similarly, the constructive interference of the MO mode in the transverse $xy$ plane cannot be solely explained by the symmetry of the $H_z$ component in Fig. \ref{F2}(e).

A further significant observation drawn from the far-field distributions in Figs. \ref{F2}(d) and \ref{F2}(e) is the absence of radiation in the $z$ direction, which can be attributed to the vectorial characteristics of the longitudinal $m_z$ dipole and its associated excited multipole modes. Understanding this vectorial nature of the electromagnetic field is crucial for our later discussion on at-$\Gamma$ BICs.

\subsection{Antibonding and bonding magnetic quadrupole modes in diatomic metamolecules}
\begin{figure}[htbp]
\includegraphics[width = \columnwidth]{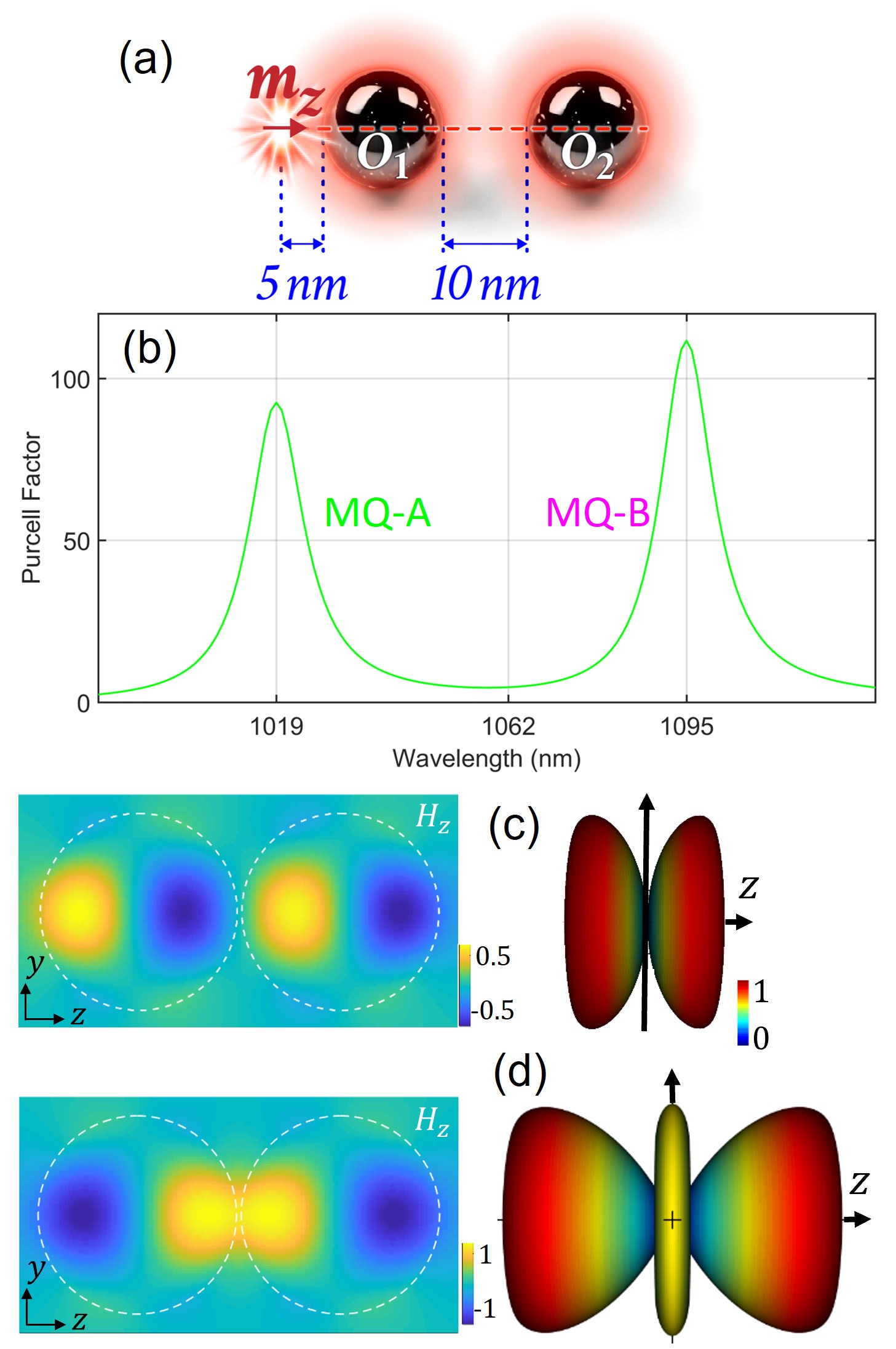}
\caption{\label{F3} (a) Interaction schematic of a $z$-oriented magnetic dipole with a diatomic photonic molecule. (b) The spectral profile of the Purcell factor illustrates the splitting of the MQ mode into antibonding (MQ-A) and bonding (MQ-B) modes. (c)-(d) Near- and far-field distributions of the collective MQ-A and MQ-B modes, respectively.}
\end{figure}

Figure \ref{F3} illustrates the splitting of the intrinsic MQ mode into antibonding (MQ-A) and bonding (MQ-B) modes upon optical coupling of the two meta-atoms, as shown schematically in Fig. \ref{F3}(a). The $F_P$ plot in Fig. \ref{F3}(b) reveals the typical blue and red shifts of the antibonding and bonding modes, akin to observations in molecular physics.

The near- and far-field distributions of the MQ-A and MQ-B modes are depicted in Figs. \ref{F3}(c) and \ref{F3}(d), respectively. Notably, the near-field $H_z$ distribution of the antibonding mode displays an anti-symmetric nature, contrasting with its bonding counterpart. The bonding nature of the MQ-B mode is evident with a magnetic hotspot at the gap between the two meta-atoms, enhancing the emission rate of magnetic emitters.

Another significant feature is the differing interference characteristics of the two modes when observing the far-field patterns in the $xy$ plane. Despite the similar $H_z$ symmetry in the meta-atoms for both modes, the transverse far fields exhibit destructive and constructive interference for the MQ-A and MQ-B modes, respectively. This observation underscores the necessity of considering collective behaviors to understand the far-field patterns. In essence, examining near-field distributions in individual units alone proves insufficient for explaining the far-field characteristics of the interaction between light emitters and metastructures.

\section{Collective nature of bound states in the continuum}
This section investigates the physics governing high-Q resonances within the frameworks of Bloch waves in photonic crystals and scattering waves in metastructures, particularly focusing on elucidating the physical mechanisms behind BICs, which conventionally associated with photonic crystals formed by an infinite periodic arrangement of spheres.

Despite significant advances in both applications and fundamental studies of BICs, their unified physical mechanism remains an open question~\cite{kang2023applications,sun2024high}. Here, we explore the two most common mechanisms underlying the existence of BICs: symmetry mismatch and the destructive interference of multipolar modes. Throughout our discussion, we illustrate how collective resonances emerging from strong multiple scattering provide a unified understanding of these high-Q  and infinite-Q resonances.

\subsection{At-$\Gamma$ bound states in the continuum and the divergence of collective resonances}

\begin{figure*}
\includegraphics[width = 16 cm]{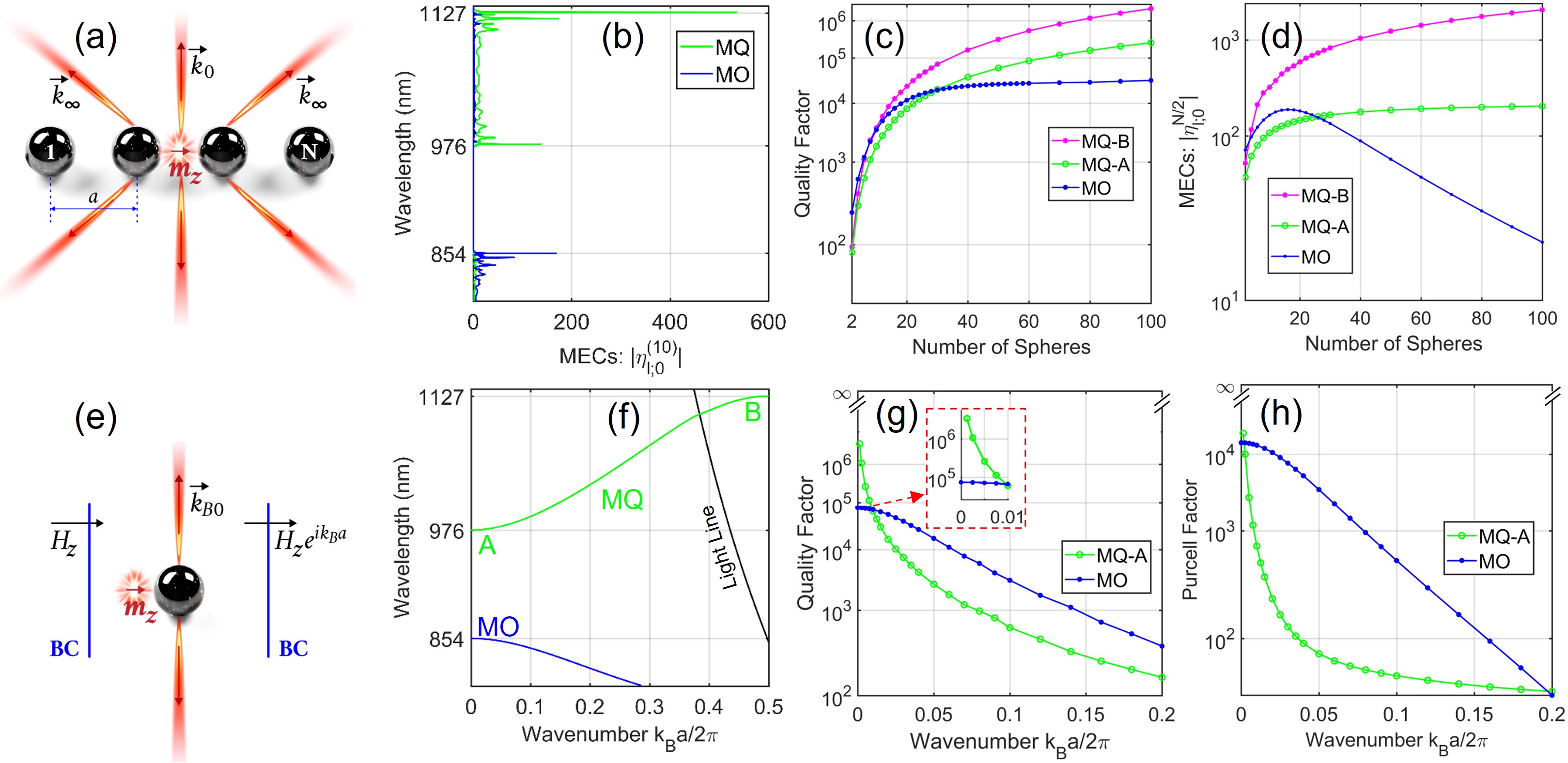}
\caption{\label{F4} Comparison of results obtained from metastructure (a-d) and photonic crystal (e-h) analyses. (a) Schematic illustrating the interaction between a magnetic dipole and a metastructure composed of $N$ spheres, within a 3D environment supporting an infinite number of degrees of freedom represented by the wave vector $\hat{k}_\infty$. (b) Spectral profiles displaying MQ and MO expansion coefficients for a 20-sphere chain, revealing two photonic bands signify the existence of two Bloch MQ and MO waves, respectively. (c) Analysis of the Q factors of MQ and MO bandedge modes, showcasing their divergence and convergence with increasing numbers of spheres. (d) Investigation into the scattering multipole coefficients representing the internal field localized at the middle sphere of the finite extent chain. (e) Schematic depicting the interaction between a magnetic dipole and a Bloch wave within the photonic crystal framework, featuring a single continuum diffraction channel denoted by the wave vector $\hat{k}_{B0}$ for our subwavelength period chain. (f) Band diagram confirming the existence of the two photonic crystal bands, with the MQ band spanning across the light line and the MO band positioned entirely above it. (g) Analysis of the Q factors associated with the two modes above the light line, indicating the existence of a bound state in the continuum corresponding to the MQ-A mode at the $\Gamma$ point. Inset highlights the divergence and convergence near the $\Gamma$ point ($k_B=0$). (h) Investigation into the Purcell factors, revealing characteristics corresponding to the Q factor analysis presented in (g).}
\end{figure*} 

Traditionally, at-$\Gamma$ BICs have been attributed to the symmetry properties of their corresponding eigenmodes, hence also termed symmetry-protected BICs. In Fig. \ref{F4}, we present an alternative perspective on the origin of at-$\Gamma$ BICs, focusing on collective resonances. Figures \ref{F4}(a) and \ref{F4}(e) depict schematics representing the two alternative views of BICs based on scattering and Bloch waves, respectively. The fundamental distinction lies in the number of radiative channels: scattering waves couple with an infinite number of plane-wave channels, whereas photonic crystals, characterized by subwavelength periods, typically feature only a single diffraction channel serving as the continuum channel for Bloch waves.

Figure \ref{F4}(b) illustrates the emergence of two distinct bands from the MQ and MO modes as we transition from 2 spheres (depicted in Fig. \ref{F3}) to 20, forming a 1D linear chain. These collective bands suggest the existence of photonic crystal bands, a hypothesis confirmed by Bloch band simulations in Fig. \ref{F4}(f). Remarkably, the wavelengths at the band edges in Fig. \ref{F4}(f) closely match predictions from MST calculations in Fig. \ref{F4}(b). Notably, the MST reveals collective resonances spread across the entire widths of the Bloch bands, both below and above the light line. This finding underscores the limited impact of the light line on cavity design using photonic metastructures. While collective resonances below the light line exhibit significantly higher strengths compared to those above it, their shared origin in resonant multiple scattering provides a unified perspective on guided resonances and Bloch bound modes.

Although the Q factors of the MQ-A and MQ-B modes for the two spheres exhibit similarity, as depicted in Fig. \ref{F3}(b) and presented in Fig. \ref{F4}(c), their behaviors diverge distinctly as the number of spheres increases. Specifically, the Q factor of the bonding MQ mode shows a notably faster increase compared to its antibonding counterpart. This divergence generally follows a power law scaling represented as $Q(N) \approx Q_0 N^{\alpha}$, where $N$ is the number of spheres. Here, we estimate $\alpha \approx 3$ for the MQ-B modes, $\alpha \approx 2$ for the MQ-A mode, and $\alpha \approx 0$ for the MO mode.

The MQ-B mode, existing below the light line, is conventionally labeled as a guided mode binding to the 1D photonic crystal without extending transversely into the far-field region. However, within  the MST framework of metacrystals, the MQ-B mode coexists with the far field via scattering effects. While modes below the light line often exhibit higher Q factors, recent attention has shifted towards guided resonances above the light line, promising infinite-Q factors for enhancing light-matter interactions. Figure \ref{F4}(g) illustrates the typical Q factor characteristics of two guided resonances corresponding to the MQ-A and MO bands. As the Bloch wavenumber approaches the $\Gamma$ point ($k_B=0$), the Q factor of the MQ-A guided resonance diverges, commonly attributed to the antisymmetric $H_z$ magnetic field component characteristic of the MQ mode, as depicted in Fig. \ref{F2}(d). Conversely, the finite $Q$ factor of the MO band arises from the symmetric $H_z$ characteristic of the MO mode, shown in Fig. \ref{F2}(e)~\cite{sidorenko2021observation}. However, challenges emerge when considering the symmetry properties of both the electric and magnetic components of these modes, as discussed earlier. The finite Q factor of the Bloch MO mode agrees with the converging Q factor obtained from the MST, as shown in Fig. \ref{F4}(c). This consistency between the MST and photonic crystal is intriguing, especially considering their distinct boundary conditions. However, it is also expected, as both arise from the Maxwell equations.

The behavior of the MQ-A mode diverging and the MO mode converging as the number of spheres increases can be mathematically likened to the divergence of the series $\sum_{n=1}^{\infty}n$ and the convergence of the series $\sum_{n=1}^{\infty}(1/(n(n+1)))$. In this analogy, the MQ mode, with its stronger long-range interactions, exhibits a strong (divergent) collective resonance, while the $Q$ factor of the MO mode remains finite. 

Further insights into the impact of the long-range interactions on the resonant MECs are elucidated in Fig. \ref{F4}(d), illustrating the divergence and convergence of the MECs representing the resonant field inside the middle sphere. Due to the positive energy of light, adding more spheres to the chain introduces both channels that enhance resonances and radiative channels that weaken them. For both the MQ-A and MQ-B modes, the enhancement effect outweighs the dissipative effect, resulting in the divergence of the MQ coefficients. Conversely, for the MO mode, the opposite occurs, leading to the convergence of the MO coefficient. These patterns of divergence and convergence in the resonant strengths elucidate the corresponding diverging and converging characteristics of the Purcell factors shown in Fig. \ref{F4}(h), where we examine the interaction between the magnetic dipole and the Bloch MQ-A and MO modes.

Interestingly, arrays of single high-Q Mie resonators (MOs) do not inherently yield higher collective resonance Q factors compared to arrays of low-Q Mie modes (MQs) with a sufficiently high number of resonators, as exemplified in Fig. \ref{F4}(c). This peculiarity arises from the intricate interplay between the behavior of individual resonators and their collective response. Individual high-Q modes retain light for longer periods but also introduce heightened radiative loss channels, limiting their collective resonances from reaching the strong multiple scattering regime necessary for divergence. In fact, higher-order Mie modes behave akin to whispering-gallery modes, and coupling these high-Q modes generally results in collective Q factors lower than those of their isolated counterparts~\cite{hoang2017fano}. This finding serves as a valuable guideline for designing coupled cavity arrays for applications such as quantum simulators and networks~\cite{hartmann2006strongly,reiserer2015cavity,chang2018colloquium}.

\begin{figure*}[htbp]
\includegraphics[width = 16 cm]{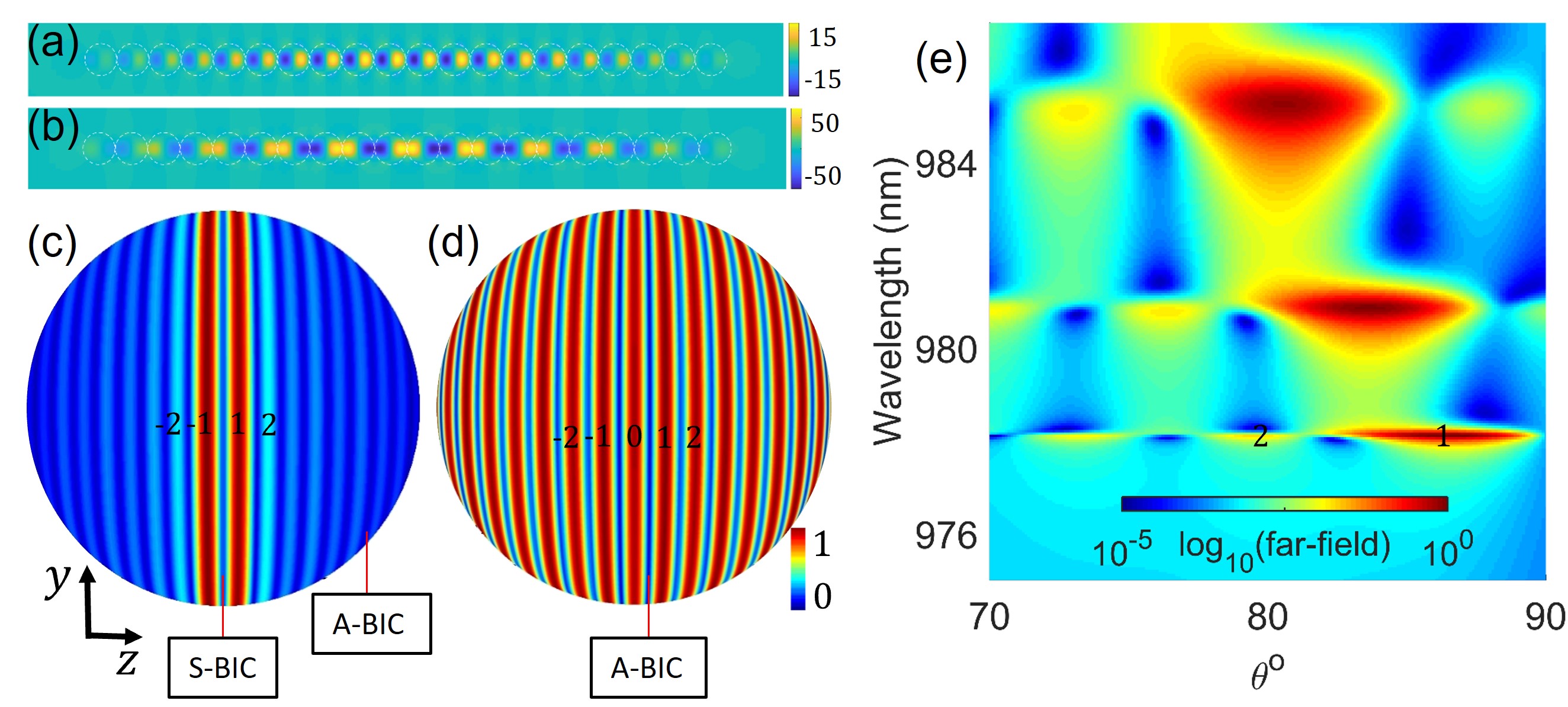}
\caption{\label{F5} Examining multipolar interference as the origin of symmetry-protected (S-BICs) and accidental (A-BICs) bound states in the continuum. Near-field $H_z$ distributions of the antibonding MQ-A and bonding MQ-B modes are depicted in (a) and (b), respectively, for the 20-sphere chain. The same set of multipoles is utilized for both modes, but the amplitude and phase differences of these multipoles result in distinct far-field interference patterns shown in (c) and (d) for the MQ-A and MQ-B modes, respectively. Panel (e) presents the far-field intensity of the antibonding mode as a function of radiation wavelength and polar angle. It's noteworthy that the wavelength and radiation angle $\theta$ of the minimum intensity differ from those associated with the at-$\Gamma$ BIC presented in Fig. \ref{F4}, which corresponds to 976 nm and 90$^{\circ}$, respectively.}
\end{figure*}

Another prominent wave mechanism proposed to explain origins of Bloch BICs is based on the destructive interference of multipole fields in specific directions \cite{sadrieva2019multipolar}. This explanation of origins of symmetry-protected (S-BIC) and accidental (A-BIC) BICs, as labeled in Fig. \ref{F5}, is rooted in coupled mode theory. However, it's essential to note a common misconception originating from this theory, where the Bloch diffraction channel with the $\overrightarrow{k}_{B0}$ wavevector is often confused with the radiation transverse $\overrightarrow{k}_0$ direction, as depicted in Fig. \ref{F4}. Additionally, at-$\Gamma$ BICs are sometimes incorrectly labeled as a nonradiative mode, implying that the amplitude of the $\overrightarrow{k}_{B0}$ diffraction wave is zero. Contrary to this misconception, our observations in Fig. \ref{F4} reveal that the amplitude of the $\overrightarrow{k}_{B0}$ diffraction wave is not zero due to the direction of the excitation magnetic dipole and its associated multipole modes supported by the unit cell.

The Q factor divergence at the $\Gamma$ point is primarily a result of enforcing the Bloch boundary conditions along the $z$ direction, which represents the long-range couplings in the infinite photonic crystal. In our FDTD simulations, this enforcement injects energy into the unit cell while allowing energy to escape along the transverse $\overrightarrow{k}_{B0}$ direction, indicating that its amplitude is not necessarily zero. At the $\Gamma$ point, the injected energy balances with the diffracted energy, resulting in the corresponding Bloch eigenmode having a real eigenvalue, often referred to as a BIC~\cite{hsu2016bound}. This non-vanishing $\overrightarrow{k}_{B0}$ mode with a real eigenvalue (infinite Q factor) aligns well with the collective resonance picture presented above.

Nevertheless, multipolar interference and coupling significantly influence the light properties in metastructures, giving rise to far-field patterns with intriguing topological characteristics. We delve deeper into these multipolar wave phenomena by examining the near- and far-field distributions of the MQ-A and MQ-B modes supported by the 20-sphere chain in Fig. \ref{F5}. Notably, despite earlier concerns regarding symmetry arguments, they offer valuable insights into the appearance of dark and bright fringes on the transverse $xy$ plane in Figs. \ref{F5}(c) and \ref{F5}(d). The antisymmetry of the antibonding mode (Fig. \ref{F5}(a)) corresponds to the dark fringe in Fig. \ref{F5}(c), while the symmetry of the bonding mode (Fig. \ref{F5}(b)) accounts for the 0th scattering order in Fig. \ref{F5}(d). However, both antibonding and bonding modes exhibit similar far-field patterns, with dark fringes positioned between bright fringes representing scattering orders. Each dark fringe results from the destructive interference of excited multipole modes within the chain. In fact, these dark fringes are ubiquitous, even appearing in the single-sphere case (Fig. \ref{F2}), concluding that these directions of minimal radiation intensities should not be considered as representations of Bloch BICs.

The analysis of the topological properties of BICs typically involves a far-field perspective, where the polar angle $\theta$ is commonly associated with the light's momentum \cite{zhen2014topological,jin2019topologically}. To deepen our understanding of far-field characteristics, we present the far-field intensity as a function of both the excitation wavelength and polar angle $\theta$ in Fig. \ref{F5}(e), focusing specifically on the antibonding MQ-A mode. It's important to note the distinction between the far-field plots in Figs. \ref{F5}(c) and \ref{F5}(e): the former represents a single wavelength across the entire polar angle range of $\theta$ ($0,\pi$) and the half azimuthal angle range $\phi$ ($0,\pi$), while the latter captures ranges of the excitation wavelength and polar angle. Due to the axial symmetry of our system, the far-field distributions are azimuthally invariant. The 1st and 2nd scattering orders from Fig. \ref{F5}(c) are marked in Fig. \ref{F5}(e) for better visualization. As explained earlier, the intensity at $\theta=90^\circ$, corresponding to the marked S-BIC in Fig. \ref{F5}(c), is not at a minimum. Instead, multiple minimal far-field points occur at various $\theta$ angles and wavelengths, some of which correspond to off-$\Gamma$ points in Fig. \ref{F4}(f), indicating guided resonances with finite $Q$ factors. Consequently, the origins of BICs relying on the destructive far-field interference of multipole modes are not universally conclusive.

\begin{figure}[htbp]
\includegraphics[width = \columnwidth]{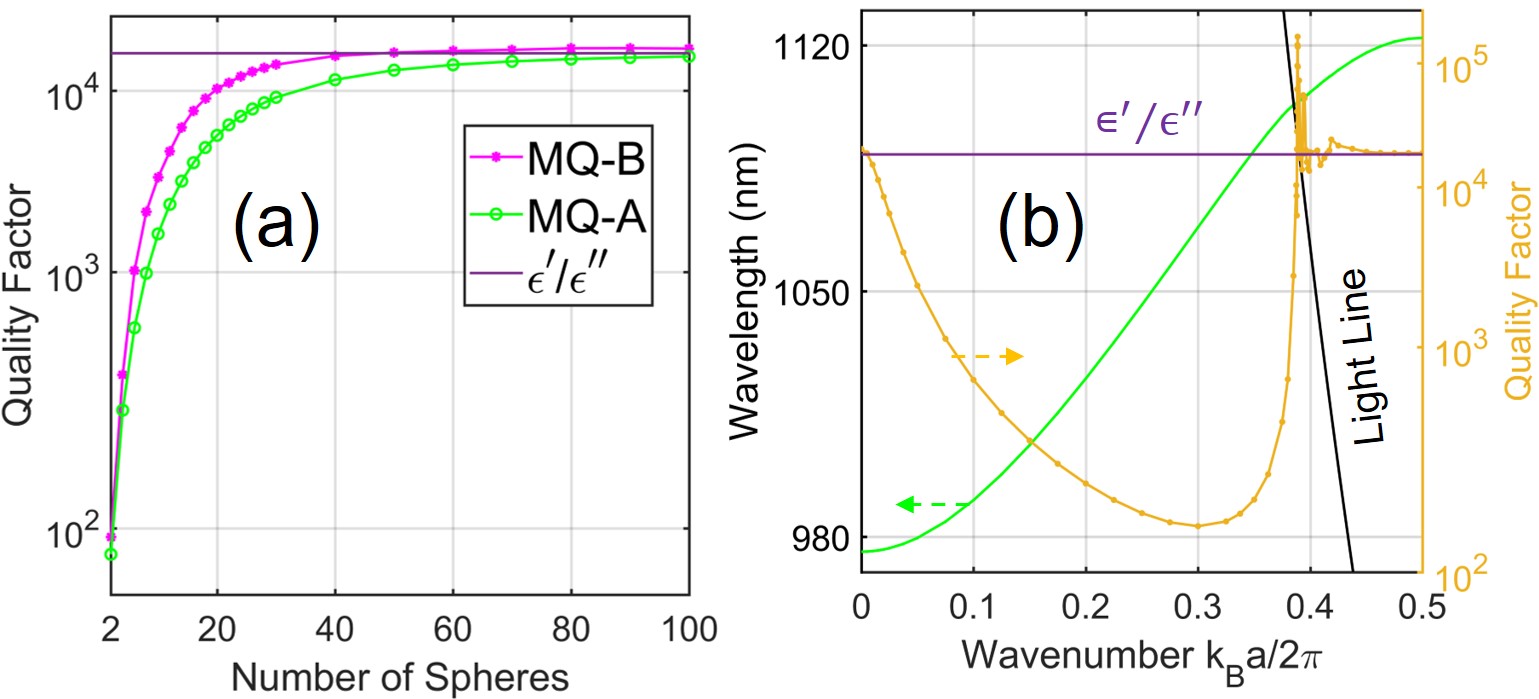}
\caption{\label{F6} Impact of material absorption loss on collective resonances and Bloch modes. (a) Both bonding and antibonding MQ modes exhibit Q factors converging to values limited by absorption loss. (b) The convergence of Q factors for the two bandedge modes corroborates findings from multiple scattering theory in (a).}
\end{figure}

The Q factor divergences illustrated in Fig. \ref{F4} pertain to chains of lossless silicon spheres. Recent research has shown a growing interest in exploring the fundamental limits of Q factors attributed to material absorption loss \cite{ustimenko2024resonances}. Typically, the Q factor is sensitive to both radiative ($Q_r$) and absorptive ($Q_{\text{abs}}$) factors, related by the equation $1/Q = 1/Q_r + 1/Q_{\text{abs}}$. For a material characterized by permittivity $\epsilon^\prime + i\epsilon^{\prime\prime}$, the absorptive Q factor is represented as $Q_{\text{abs}} = \epsilon^\prime/\epsilon^{\prime\prime}$.

In Fig. \ref{F6}, we analyze the Q factor characteristics of MQ-A and MQ-B modes for a lossy silicon material with a refractive index of $3.5 + i10^{-4}$, corresponding to $Q_{\text{abs}} = 1.75 \times 10^4$. In contrast to the rapid divergence of Q factors for MQ-A and MQ-B modes in the absence of absorption loss, Fig. \ref{F6}(a) reveals their convergence to approximately the $Q_{\text{abs}}$ value in the presence of absorption loss, highlighting the severity of this limitation. The observation that saturated Q factors exceed $Q_{\text{abs}}$ can be attributed to the interplay between long-range couplings in the chains and absorption loss. Additionally, these saturated Q factors are influenced by the fraction of electromagnetic energy stored inside the spheres \cite{mikhailovskii2024engineering}.

Furthermore, Fig. \ref{F6}(b) uncovers an intriguing finding: the maximum Q factor of Bloch modes surpasses $Q_{\text{abs}}$ by an order of magnitude and differs from the bandedge MQ-B mode predicted by the MST. Instead, it corresponds to the Bloch mode near the light line. This discrepancy arises from the tighter confinement of the Bloch bandedge mode to the spheres, resulting in increased absorption loss. However, it's crucial to note that for finite chains, the Q factor of the bandedge MQ-B mode consistently remains the maximum, emphasizing the importance of accounting for practical finite sizes in light-matter interaction systems.

\subsection{Off-$\Gamma$ bound states in the continuum}

\begin{figure*}[htbp]
\includegraphics[width = 16 cm]{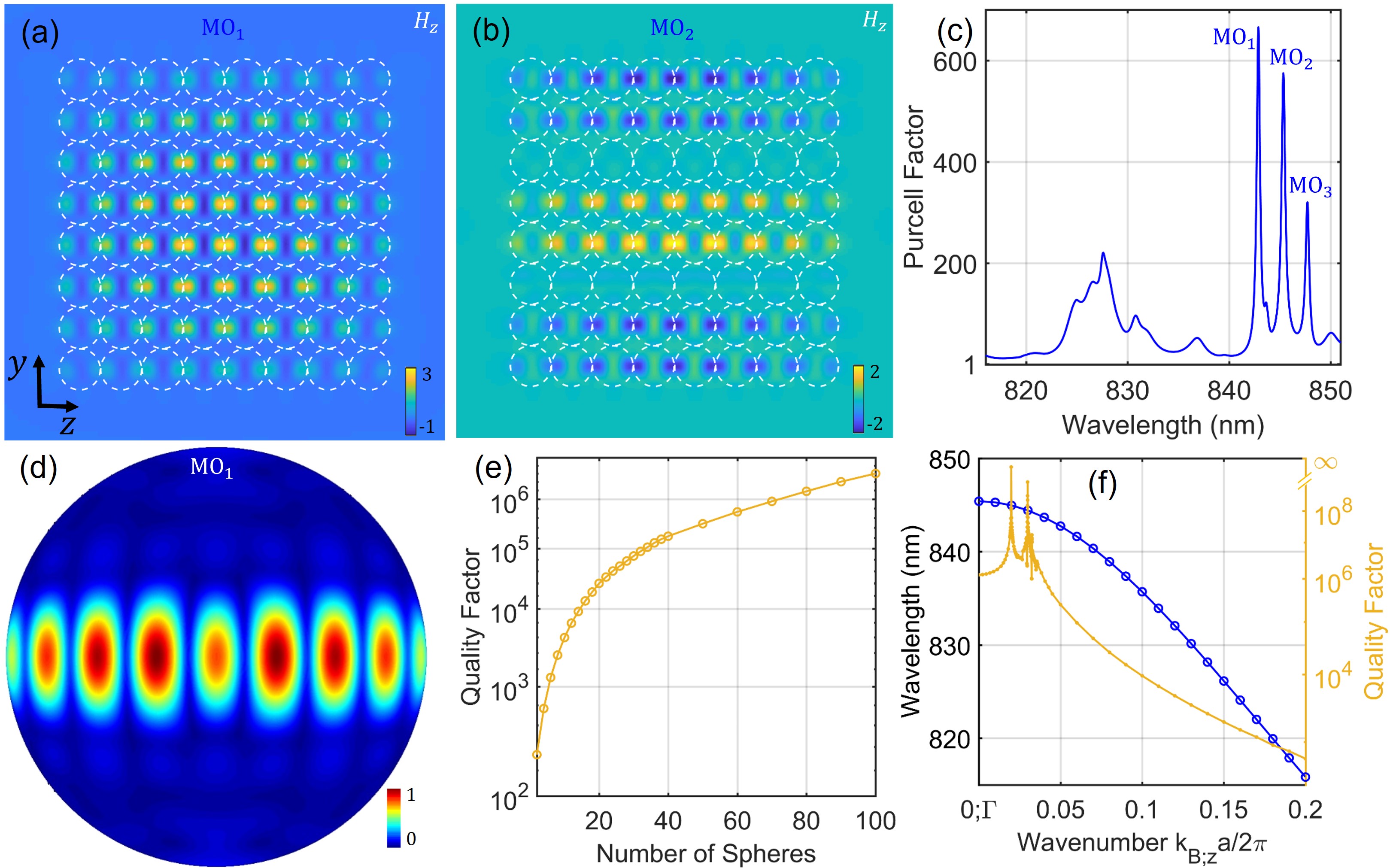}
\caption{\label{F7} Off-$\Gamma$ bound states in the continuum from the collective MO resonances in square arrays of spheres. (a)-(b) Near-field $H_z$ distributions corresponding to the $\text{MO}_1$ and $\text{MO}_2$ modes marked in (c), where we show the Purcell factor associated with the magnetic dipole placed at the strongest field in (a). (d) Far-field distribution of the $\text{MO}_1$ mode. (e) Divergence of the Q factor of the $\text{MO}_1$ mode with an increasing number of spheres, representing the size in one dimension of the square array. (f) Band structure and divergence of the Q factor at off-$\Gamma$ points indicate the existence of accidental bound states in the continuum.}
\end{figure*}

This subsection presents the collective resonance origin of off-$\Gamma$ BICs. Traditionally, the existence of these off-$\Gamma$ BICs is challenging to predict, hence termed as accidental BICs~\cite{sidorenko2021observation}. As demonstrated in the preceding subsection, the at-$\Gamma$ BIC arises from divergence of sum of the resonant scattering multipole fields, while the convergence of the Q factor of the MO mode in the 1D photonic crystal is a result of radiative loss overwhelming the resonant enhancement due to the long-range coupling effect. An effective strategy to mitigate this radiative loss is to augment the structure by adding more spheres to form a 2D array or a metasurface, as illustrated in Fig. \ref{F7}. Transforming the 1D chain into the 2D metastructure allows for the scattering of light escaping in the $y$ direction back towards the central region, thereby enhancing the light trapping efficiency.

Resonant metasurfaces typically accommodate numerous supermodes~\cite{overvig2020selection,geints2023phase,geints2023manipulating}, as represented by MO$_{1,2,3}$ in Figs. \ref{F7}(a)-(c). The near-field $H_z$ distributions of the two MO$_1$ and MO$_2$ supermodes clearly show their collective characteristics. Remarkably, we observe that the dominant MO$_1$ supermode differs from the bandedge nature observed in its 1D chain counterpart (Fig. \ref{F4}). This observation suggests that the maximal Q factor of the corresponding guided resonance occurs not at the $\Gamma$ point but at an off-$\Gamma$ point. The far-field distribution of the MO$_1$ supermode shown in Fig. \ref{F7}(d) has several null-field directions, a common feature of collective resonances.    

Figure \ref{F7}(e) presents the divergence of the Q factor of the MO$_1$ supermode with an increasing number of spheres forming square metastructures. This Q factor divergence suggests that we could observe BICs with the corresponding 2D photonic crystals, considering the divergence of Q factors of collective resonances as a mechanism for detecting BICs. This hypothesis finds support in simulations of the photonic crystal band and its associated Q factor as depicted in Fig. \ref{F7}(f). Our findings shed light on an intriguing aspect: while the Q factors of BICs theoretically tend to infinity, experimental results commonly cap at levels less than one million~\cite{jin2019topologically}. As illustrated in Fig. \ref{F7}(e), even when scaling the structure to a $100\times100$ resonator array, the Q factor approaches only around 1 million. Notably, a 2D supermode supported by a $16\times16$ array of MOs was employed to realize a BIC laser with a Q factor in the range of several thousands~\cite{kodigala2017lasing}, consistent with our Q factor simulation presented in Fig. \ref{F7}(e). By comparing these results with those obtained for the case of 1D arrays, as shown in Fig. \ref{F4}(c), the utilization of  2D arrays of MQs may significantly enhance the efficiency of MO-based surface-emitting lasers.   

Conceptually, the off-$\Gamma$ BICs depicted in Fig. \ref{F7}(f) have often been characterized as Friedrich–Wintgen BICs, attributed to their origin from coupled resonances \cite{kodigala2017lasing}. However, the existence of Friedrich–Wintgen BICs necessitates a critical condition: the number of closed channels must exceed the continuum channels \cite{hsu2016bound}. As previously discussed, this condition is not strictly met by light waves. A more apt term to describe the off-$\Gamma$ BICs might be Feshbach-type BICs, drawing on the work of Feshbach, who provided a unified theory of coupled resonances interacting with various open channels~\cite{feshbach1958unified}. Feshbach's theory shares similarities with our MST approach and is therefore better suited to explain the existence of the off-$\Gamma$ BICs.

Past efforts to explore the impact of finite size effects on off-$\Gamma$ BICs relied on the tight-binding model for Bloch waves, attributing scattering loss primarily to the edges of finite structures~\cite{bulgakov2019high,mikhailovskii2024engineering}. However, our MST offers a contrasting perspective, revealing that scattering loss predominantly originates from the central regions for both at-$\Gamma$ and off-$\Gamma$ BICs~\cite{hoang2024photonic}. These differing perspectives stem from the analogous yet distinct behaviors between matter and light waves. The concept of Bloch waves, rooted in matter waves traversing crystal lattices without scattering, hinges on two fundamental properties. Firstly, matter waves, with potential negative energies, limit interactions with the 3D environment, allowing the matter waves and their potential structures to be considered closed systems. Secondly, these matter waves, when interacting with their local lattice potentials, do not encounter a retardation effect due to their probabilistic nature~\cite{lagendijk1996resonant}.

Conversely, light waves possess positive energies, thus the internal fields within the unit cells of metacrystals--or metastructures in general--maintain coupling with the 3D environment even when we extend their structure to infinity. Moreover, in scenarios of resonant multiple scattering, light waves experience retardation effects, a crucial consideration. Unlike matter waves, the localization of light in resonant metacrystals does not rely on structural disorders, which are required for localizing matter waves through the Anderson effect. Essentially, the central localization of light within the resonant metasurface, as depicted in Fig. \ref{F7}(a), and its associated van Hove singularity at the off-$\Gamma$ points stem from the distinctive retardation effects inherent to light waves. Our findings elucidate these distinct attributes of photonic BICs, offering clarity on their physical origin and promising avenues for advancements in `Mie-tronics' studies. 

\section{Discussion and conclusion}

{\bf Near-field and far-field excitation of collective resonances.} We presented a comprehensive picture of high-Q resonances, spanning from whispering-gallery modes to collective resonances in photonic metastructures. Our focus was on resonant multiple scattering and the multipole expansion of the electromagnetic field, with the primary goal of optimizing light-matter interactions. While our investigation centered on the near-field excitation of high-Q resonances, our methodology also offers insights into far-field excitation schemes by treating the 3D environment as a practical continuum. The coupling between high-Q metastructures and incident free-space beams is by no means trivial. Exciting high-Q resonances from the far field proves challenging due to the complexity of the intrinsic multipolar content of these collective modes. Simple laser beams may prove inadequate as their intrinsic modes might not align suitably. Efficient excitation requires the use of structured light beams with appropriate multipole content~\cite{hoang2012multipole, bliokh2023roadmap}. Thus, our proposed Mie-tronics approach not only aids in developing effective design strategies of high-Q resonances but also provides an alternative to excitation methods based on Bloch waves and group theory~\cite{overvig2020selection}.

{\bf Scaling law of collective resonances.} Our findings offer valuable insights into the fundamental quantitative limits of Q and Purcell factors within resonant metastructures. In these resonant systems of size $N$, their Q and Purcell factors follow a power law scaling $Q(N)\approx N^\alpha$. The primary objective when optimizing systems of fixed size $N$ is to maximize the $\alpha$ factor~\cite{mikhailovskii2024engineering}. This scaling parameter is influenced by various system attributes, including its refractive index, unit cell geometry, and notably, the dominant Mie modes contributing to collective resonances. We find that generally collective bonding modes exhibit a scaling factor of $\alpha=3$, at least one polynomial degree higher than its antibonding counterpart. To further augment this scaling factor, the most promising strategy involves finely tuning the geometrical parameters to merge collective resonances and achieve the so-called merging BICs or flatband resonances, as exemplified by a scaling factor of $\alpha=6$ in our recent study~\cite{hoang2024photonic}.

Interestingly, the scaling law and fine-tuning effects are also observable in collective responses of field-mediated atoms, each supporting a two-level dipolar transition state~\cite{zhang2020subradiant,volkov2024strongly}. The similarity in behavior between atomic and photonic systems arises from the analogy between a resonant mode and a two-level atomic system, as well as the significant role of multiple light scattering in both scenarios~\cite{lagendijk1996resonant,asselie2022optical}. It is noteworthy that while dipole interactions are vital, our photonic systems encompass more general interactions, including higher-order multipole modes, for a comprehensive understanding of collective responses.

{\bf Similarities and differences between photonic and matter BICs.} Conceptually, our results provide a clear picture of photonic and matter BICs, highlighting both their similarities and differences. The original BIC concept revolves around electron localization in potentials extending infinitely, which supports the BIC as an infinitely narrow resonance~\cite{stillinger1975bound}. Any introduction of finiteness to such potentials transforms the electronic BIC into a finite-Q resonance~\cite{capasso1992observation}. This original BIC is characterized by the divergence of the sum of partial scattering waves~\cite{stillinger1976potentials}. Our understanding of BICs shares two key similarities: our resonant structures also demand infinite extension to induce the divergence of partial multipole waves, and introducing 3D finiteness causes the transition of infinitely narrow resonances into finite-Q resonances.

However, a difference exists between our photonic BICs and the original matter wave BIC. While our photonic BICs maintain coupling to the 3D environment, the original matter wave BIC primarily focused on 1D matter waves. It employed layered potential structures that neglect coupling to additional dimensions, consequently considering a single continuum channel only. The crucial distinction for our photonic BICs, existing despite their coupling to the far-field region, lies in the impact of retardation effects from Mie resonances. These effects temporarily confine light waves within resonators, facilitating strong multiple back-scattering and resulting in the divergence of partial multipole waves at the center of photonic structures. This divergence means that achieving high-Q resonances involves maximizing the scattering multipole coefficients, leading to the appearance of numerous radiation fringes in the far field. Interestingly, the oscillating far-field characteristic of our photonic BICs is also a feature of compact matter BICs, also known as Friedrich–Wintgen BICs~\cite{fonda1963bound,friedrich1985interfering}. However, the existence of these compact matter BICs hinges on closed channels associated with negative-energy states—an aspect not applicable to light waves, which inherently possess positive energies. Consequently, all light modes operate as open channels, precluding the feasibility of compact photonic BICs.

{\bf Collective nature as a unified picture of photonic BICs.} The origin of photonic BICs traces back to the analogy drawn between resonances and closed channels, alongside the association between diffraction orders and open channels~\cite{marinica2008bound}. Subsequently, photonic BICs has garnered significant interest from the photonic crystal community due to both the familiarity with Bloch waves and their potential applications across diverse fields. BIC research has often focused on bands above the light line in the band diagram~\cite{hsu2013observation, zhen2014topological, hsu2016bound, jin2019topologically}, associating BICs with isolated Bloch wavenumbers ($k_\text{B}$) and infinite Q factors, suggesting their decoupling from the far-field region~\cite{dong2022nanoscale,sun2024high}.

However, our findings challenge this understanding, revealing a paradigm shift: BICs correspond to the divergence of collective resonances coupling to the far-field region, even for infinitely extended metastructures. Consequently, the widely held belief that BICs collapse in the presence of fabrication defects and structural disorders should be reconsidered~\cite{xu2023recent}. The collective nature of BIC resonances makes them robust to fabrication defects and structural disorders. Constituent resonators adapt in phase to structural variations to maintain their collective responses. Moderate structural variations may shift BIC resonances in spectral space, affecting their strength but not destroying them. Our identification of the collective nature of BICs explains their robust existence, an alternative to their commonly associated topological nature~\cite{zhen2014topological,jin2019topologically}. It's worth mentioning that multipoles inherently possess topological properties, providing an explanation for the topological nature of BICs. This reevaluation of the origin of infinite Q factors fundamentally reshapes our understanding of BICs, revealing the intricate interplay between resonances, scattering, and the far-field region.

In conclusion, the allure of high-$Q$ resonances lies in their potential to facilitate a robust interaction between light and matter. Our research underscores the significant promise of bonding resonances as a pathway to achieving substantial enhancements in interaction efficiency. The plethora of collective resonances may empower photonic metastructures with the capacity to improve interaction efficiencies across a wide spectral range, encompassing linear to nonlinear optics, from THz to visible light, and beyond. This study highlights the tremendous potential of Mie-tronics in applications reliant on high-$Q$ resonances, thereby advancing our comprehension of the BIC phenomena, and their prospective impact on photonics.

\begin{acknowledgments}

This research has been supported by the Agency for Science, Technology and Research (A$^\star$STAR) under its Career Development Fund (C210112012). D.L. acknowledges a support from the National Research Foundation, Singapore and A*STAR under its CQT Bridging Grant. Y.K. was supported by the Australian Research Council (Grant DP210101292) and the International Technology Center Indo-Pacific (ITC IPAC) via Army Research Office (contract FA520923C0023).

\end{acknowledgments}

\bibliography{Reference}

\begin{thebibliography}{55}%
\makeatletter
\providecommand \@ifxundefined [1]{%
 \@ifx{#1\undefined}
}%
\providecommand \@ifnum [1]{%
 \ifnum #1\expandafter \@firstoftwo
 \else \expandafter \@secondoftwo
 \fi
}%
\providecommand \@ifx [1]{%
 \ifx #1\expandafter \@firstoftwo
 \else \expandafter \@secondoftwo
 \fi
}%
\providecommand \natexlab [1]{#1}%
\providecommand \enquote  [1]{``#1''}%
\providecommand \bibnamefont  [1]{#1}%
\providecommand \bibfnamefont [1]{#1}%
\providecommand \citenamefont [1]{#1}%
\providecommand \href@noop [0]{\@secondoftwo}%
\providecommand \href [0]{\begingroup \@sanitize@url \@href}%
\providecommand \@href[1]{\@@startlink{#1}\@@href}%
\providecommand \@@href[1]{\endgroup#1\@@endlink}%
\providecommand \@sanitize@url [0]{\catcode `\\12\catcode `\$12\catcode `\&12\catcode `\#12\catcode `\^12\catcode `\_12\catcode `\%12\relax}%
\providecommand \@@startlink[1]{}%
\providecommand \@@endlink[0]{}%
\providecommand \url  [0]{\begingroup\@sanitize@url \@url }%
\providecommand \@url [1]{\endgroup\@href {#1}{\urlprefix }}%
\providecommand \urlprefix  [0]{URL }%
\providecommand \Eprint [0]{\href }%
\providecommand \doibase [0]{https://doi.org/}%
\providecommand \selectlanguage [0]{\@gobble}%
\providecommand \bibinfo  [0]{\@secondoftwo}%
\providecommand \bibfield  [0]{\@secondoftwo}%
\providecommand \translation [1]{[#1]}%
\providecommand \BibitemOpen [0]{}%
\providecommand \bibitemStop [0]{}%
\providecommand \bibitemNoStop [0]{.\EOS\space}%
\providecommand \EOS [0]{\spacefactor3000\relax}%
\providecommand \BibitemShut  [1]{\csname bibitem#1\endcsname}%
\let\auto@bib@innerbib\@empty
\bibitem [{\citenamefont {Mahmoodian}\ \emph {et~al.}(2017)\citenamefont {Mahmoodian}, \citenamefont {Prindal-Nielsen}, \citenamefont {S\"{o}llner}, \citenamefont {Stobbe},\ and\ \citenamefont {Lodahl}}]{Mahmoodian:17}%
  \BibitemOpen
  \bibfield  {author} {\bibinfo {author} {\bibfnamefont {S.}~\bibnamefont {Mahmoodian}}, \bibinfo {author} {\bibfnamefont {K.}~\bibnamefont {Prindal-Nielsen}}, \bibinfo {author} {\bibfnamefont {I.}~\bibnamefont {S\"{o}llner}}, \bibinfo {author} {\bibfnamefont {S.}~\bibnamefont {Stobbe}},\ and\ \bibinfo {author} {\bibfnamefont {P.}~\bibnamefont {Lodahl}},\ }\bibfield  {title} {\bibinfo {title} {Engineering chiral light-matter interaction in photonic crystal waveguides with slow light},\ }\href {https://doi.org/10.1364/OME.7.000043} {\bibfield  {journal} {\bibinfo  {journal} {Opt. Mater. Express}\ }\textbf {\bibinfo {volume} {7}},\ \bibinfo {pages} {43} (\bibinfo {year} {2017})}\BibitemShut {NoStop}%
\bibitem [{\citenamefont {Lu}\ \emph {et~al.}(2022)\citenamefont {Lu}, \citenamefont {McClung},\ and\ \citenamefont {Srinivasan}}]{Lu2022}%
  \BibitemOpen
  \bibfield  {author} {\bibinfo {author} {\bibfnamefont {X.}~\bibnamefont {Lu}}, \bibinfo {author} {\bibfnamefont {A.}~\bibnamefont {McClung}},\ and\ \bibinfo {author} {\bibfnamefont {K.}~\bibnamefont {Srinivasan}},\ }\bibfield  {title} {\bibinfo {title} {High-{Q} slow light and its localization in a photonic crystal microring},\ }\href {https://doi.org/10.1038/s41566-021-00912-w} {\bibfield  {journal} {\bibinfo  {journal} {Nature Photonics}\ }\textbf {\bibinfo {volume} {16}},\ \bibinfo {pages} {66} (\bibinfo {year} {2022})}\BibitemShut {NoStop}%
\bibitem [{\citenamefont {Gonz{\'a}lez-Tudela}\ \emph {et~al.}(2024)\citenamefont {Gonz{\'a}lez-Tudela}, \citenamefont {Reiserer}, \citenamefont {Garc{\'\i}a-Ripoll},\ and\ \citenamefont {Garc{\'\i}a-Vidal}}]{gonzalez2024light}%
  \BibitemOpen
  \bibfield  {author} {\bibinfo {author} {\bibfnamefont {A.}~\bibnamefont {Gonz{\'a}lez-Tudela}}, \bibinfo {author} {\bibfnamefont {A.}~\bibnamefont {Reiserer}}, \bibinfo {author} {\bibfnamefont {J.~J.}\ \bibnamefont {Garc{\'\i}a-Ripoll}},\ and\ \bibinfo {author} {\bibfnamefont {F.~J.}\ \bibnamefont {Garc{\'\i}a-Vidal}},\ }\bibfield  {title} {\bibinfo {title} {Light--matter interactions in quantum nanophotonic devices},\ }\href@noop {} {\bibfield  {journal} {\bibinfo  {journal} {Nature Reviews Physics}\ ,\ \bibinfo {pages} {1}} (\bibinfo {year} {2024})}\BibitemShut {NoStop}%
\bibitem [{\citenamefont {Kivshar}(2022)}]{kivshar2022rise}%
  \BibitemOpen
  \bibfield  {author} {\bibinfo {author} {\bibfnamefont {Y.}~\bibnamefont {Kivshar}},\ }\bibfield  {title} {\bibinfo {title} {The rise of {M}ie-tronics},\ }\href@noop {} {\bibfield  {journal} {\bibinfo  {journal} {Nano Letters}\ }\textbf {\bibinfo {volume} {22}},\ \bibinfo {pages} {3513} (\bibinfo {year} {2022})}\BibitemShut {NoStop}%
\bibitem [{\citenamefont {Shastri}\ and\ \citenamefont {Monticone}(2023)}]{shastri2023nonlocal}%
  \BibitemOpen
  \bibfield  {author} {\bibinfo {author} {\bibfnamefont {K.}~\bibnamefont {Shastri}}\ and\ \bibinfo {author} {\bibfnamefont {F.}~\bibnamefont {Monticone}},\ }\bibfield  {title} {\bibinfo {title} {Nonlocal flat optics},\ }\href@noop {} {\bibfield  {journal} {\bibinfo  {journal} {Nature Photonics}\ }\textbf {\bibinfo {volume} {17}},\ \bibinfo {pages} {36} (\bibinfo {year} {2023})}\BibitemShut {NoStop}%
\bibitem [{\citenamefont {Schiattarella}\ \emph {et~al.}(2024)\citenamefont {Schiattarella}, \citenamefont {Romano}, \citenamefont {Sirleto}, \citenamefont {Mocella}, \citenamefont {Rendina}, \citenamefont {Lanzio}, \citenamefont {Riminucci}, \citenamefont {Schwartzberg}, \citenamefont {Cabrini}, \citenamefont {Chen} \emph {et~al.}}]{schiattarella2024directive}%
  \BibitemOpen
  \bibfield  {author} {\bibinfo {author} {\bibfnamefont {C.}~\bibnamefont {Schiattarella}}, \bibinfo {author} {\bibfnamefont {S.}~\bibnamefont {Romano}}, \bibinfo {author} {\bibfnamefont {L.}~\bibnamefont {Sirleto}}, \bibinfo {author} {\bibfnamefont {V.}~\bibnamefont {Mocella}}, \bibinfo {author} {\bibfnamefont {I.}~\bibnamefont {Rendina}}, \bibinfo {author} {\bibfnamefont {V.}~\bibnamefont {Lanzio}}, \bibinfo {author} {\bibfnamefont {F.}~\bibnamefont {Riminucci}}, \bibinfo {author} {\bibfnamefont {A.}~\bibnamefont {Schwartzberg}}, \bibinfo {author} {\bibfnamefont {S.}~\bibnamefont {Cabrini}}, \bibinfo {author} {\bibfnamefont {J.}~\bibnamefont {Chen}}, \emph {et~al.},\ }\bibfield  {title} {\bibinfo {title} {Directive giant upconversion by supercritical bound states in the continuum},\ }\href@noop {} {\bibfield  {journal} {\bibinfo  {journal} {Nature}\ }\textbf {\bibinfo {volume} {626}},\ \bibinfo {pages} {765} (\bibinfo {year} {2024})}\BibitemShut {NoStop}%
\bibitem [{\citenamefont {Hsu}\ \emph {et~al.}(2016)\citenamefont {Hsu}, \citenamefont {Zhen}, \citenamefont {Stone}, \citenamefont {Joannopoulos},\ and\ \citenamefont {Solja{\v{c}}i{\'c}}}]{hsu2016bound}%
  \BibitemOpen
  \bibfield  {author} {\bibinfo {author} {\bibfnamefont {C.~W.}\ \bibnamefont {Hsu}}, \bibinfo {author} {\bibfnamefont {B.}~\bibnamefont {Zhen}}, \bibinfo {author} {\bibfnamefont {A.~D.}\ \bibnamefont {Stone}}, \bibinfo {author} {\bibfnamefont {J.~D.}\ \bibnamefont {Joannopoulos}},\ and\ \bibinfo {author} {\bibfnamefont {M.}~\bibnamefont {Solja{\v{c}}i{\'c}}},\ }\bibfield  {title} {\bibinfo {title} {Bound states in the continuum},\ }\href@noop {} {\bibfield  {journal} {\bibinfo  {journal} {Nature Reviews Materials}\ }\textbf {\bibinfo {volume} {1}},\ \bibinfo {pages} {1} (\bibinfo {year} {2016})}\BibitemShut {NoStop}%
\bibitem [{\citenamefont {Kang}\ \emph {et~al.}(2023)\citenamefont {Kang}, \citenamefont {Liu}, \citenamefont {Chan},\ and\ \citenamefont {Xiao}}]{kang2023applications}%
  \BibitemOpen
  \bibfield  {author} {\bibinfo {author} {\bibfnamefont {M.}~\bibnamefont {Kang}}, \bibinfo {author} {\bibfnamefont {T.}~\bibnamefont {Liu}}, \bibinfo {author} {\bibfnamefont {C.}~\bibnamefont {Chan}},\ and\ \bibinfo {author} {\bibfnamefont {M.}~\bibnamefont {Xiao}},\ }\bibfield  {title} {\bibinfo {title} {Applications of bound states in the continuum in photonics},\ }\href@noop {} {\bibfield  {journal} {\bibinfo  {journal} {Nature Reviews Physics}\ }\textbf {\bibinfo {volume} {5}},\ \bibinfo {pages} {659} (\bibinfo {year} {2023})}\BibitemShut {NoStop}%
\bibitem [{\citenamefont {Jin}\ \emph {et~al.}(2019)\citenamefont {Jin}, \citenamefont {Yin}, \citenamefont {Ni}, \citenamefont {Solja{\v{c}}i{\'c}}, \citenamefont {Zhen},\ and\ \citenamefont {Peng}}]{jin2019topologically}%
  \BibitemOpen
  \bibfield  {author} {\bibinfo {author} {\bibfnamefont {J.}~\bibnamefont {Jin}}, \bibinfo {author} {\bibfnamefont {X.}~\bibnamefont {Yin}}, \bibinfo {author} {\bibfnamefont {L.}~\bibnamefont {Ni}}, \bibinfo {author} {\bibfnamefont {M.}~\bibnamefont {Solja{\v{c}}i{\'c}}}, \bibinfo {author} {\bibfnamefont {B.}~\bibnamefont {Zhen}},\ and\ \bibinfo {author} {\bibfnamefont {C.}~\bibnamefont {Peng}},\ }\bibfield  {title} {\bibinfo {title} {Topologically enabled ultrahigh-{Q} guided resonances robust to out-of-plane scattering},\ }\href@noop {} {\bibfield  {journal} {\bibinfo  {journal} {Nature}\ }\textbf {\bibinfo {volume} {574}},\ \bibinfo {pages} {501} (\bibinfo {year} {2019})}\BibitemShut {NoStop}%
\bibitem [{\citenamefont {Joannopoulos}\ \emph {et~al.}(2008)\citenamefont {Joannopoulos}, \citenamefont {Johnson}, \citenamefont {Winn},\ and\ \citenamefont {Meade}}]{Joan2008}%
  \BibitemOpen
  \bibfield  {author} {\bibinfo {author} {\bibfnamefont {J.~D.}\ \bibnamefont {Joannopoulos}}, \bibinfo {author} {\bibfnamefont {S.~G.}\ \bibnamefont {Johnson}}, \bibinfo {author} {\bibfnamefont {J.~N.}\ \bibnamefont {Winn}},\ and\ \bibinfo {author} {\bibfnamefont {R.~D.}\ \bibnamefont {Meade}},\ }\href@noop {} {\emph {\bibinfo {title} {Photonic Crystals: Molding the Flow of Light}}}\ (\bibinfo  {publisher} {Princeton University Press},\ \bibinfo {year} {2008})\BibitemShut {NoStop}%
\bibitem [{\citenamefont {Hwang}\ \emph {et~al.}(2021)\citenamefont {Hwang}, \citenamefont {Lee}, \citenamefont {Kim}, \citenamefont {Jeong}, \citenamefont {Kwon}, \citenamefont {Koshelev}, \citenamefont {Kivshar},\ and\ \citenamefont {Park}}]{hwang2021ultralow}%
  \BibitemOpen
  \bibfield  {author} {\bibinfo {author} {\bibfnamefont {M.-S.}\ \bibnamefont {Hwang}}, \bibinfo {author} {\bibfnamefont {H.-C.}\ \bibnamefont {Lee}}, \bibinfo {author} {\bibfnamefont {K.-H.}\ \bibnamefont {Kim}}, \bibinfo {author} {\bibfnamefont {K.-Y.}\ \bibnamefont {Jeong}}, \bibinfo {author} {\bibfnamefont {S.-H.}\ \bibnamefont {Kwon}}, \bibinfo {author} {\bibfnamefont {K.}~\bibnamefont {Koshelev}}, \bibinfo {author} {\bibfnamefont {Y.}~\bibnamefont {Kivshar}},\ and\ \bibinfo {author} {\bibfnamefont {H.-G.}\ \bibnamefont {Park}},\ }\bibfield  {title} {\bibinfo {title} {Ultralow-threshold laser using super-bound states in the continuum},\ }\href@noop {} {\bibfield  {journal} {\bibinfo  {journal} {Nature Communications}\ }\textbf {\bibinfo {volume} {12}},\ \bibinfo {pages} {4135} (\bibinfo {year} {2021})}\BibitemShut {NoStop}%
\bibitem [{\citenamefont {Chen}\ \emph {et~al.}(2022)\citenamefont {Chen}, \citenamefont {Yin}, \citenamefont {Jin}, \citenamefont {Zheng}, \citenamefont {Zhang}, \citenamefont {Wang}, \citenamefont {He}, \citenamefont {Zhen},\ and\ \citenamefont {Peng}}]{chen2022observation}%
  \BibitemOpen
  \bibfield  {author} {\bibinfo {author} {\bibfnamefont {Z.}~\bibnamefont {Chen}}, \bibinfo {author} {\bibfnamefont {X.}~\bibnamefont {Yin}}, \bibinfo {author} {\bibfnamefont {J.}~\bibnamefont {Jin}}, \bibinfo {author} {\bibfnamefont {Z.}~\bibnamefont {Zheng}}, \bibinfo {author} {\bibfnamefont {Z.}~\bibnamefont {Zhang}}, \bibinfo {author} {\bibfnamefont {F.}~\bibnamefont {Wang}}, \bibinfo {author} {\bibfnamefont {L.}~\bibnamefont {He}}, \bibinfo {author} {\bibfnamefont {B.}~\bibnamefont {Zhen}},\ and\ \bibinfo {author} {\bibfnamefont {C.}~\bibnamefont {Peng}},\ }\bibfield  {title} {\bibinfo {title} {Observation of miniaturized bound states in the continuum with ultra-high quality factors},\ }\href@noop {} {\bibfield  {journal} {\bibinfo  {journal} {Science Bulletin}\ }\textbf {\bibinfo {volume} {67}},\ \bibinfo {pages} {359} (\bibinfo {year} {2022})}\BibitemShut {NoStop}%
\bibitem [{\citenamefont {Wu}\ \emph {et~al.}(2020)\citenamefont {Wu}, \citenamefont {Wang}, \citenamefont {Yang}, \citenamefont {Ji}, \citenamefont {Shen}, \citenamefont {Bao}, \citenamefont {Gao},\ and\ \citenamefont {Vahala}}]{Wu:20}%
  \BibitemOpen
  \bibfield  {author} {\bibinfo {author} {\bibfnamefont {L.}~\bibnamefont {Wu}}, \bibinfo {author} {\bibfnamefont {H.}~\bibnamefont {Wang}}, \bibinfo {author} {\bibfnamefont {Q.}~\bibnamefont {Yang}}, \bibinfo {author} {\bibfnamefont {Q.-X.}\ \bibnamefont {Ji}}, \bibinfo {author} {\bibfnamefont {B.}~\bibnamefont {Shen}}, \bibinfo {author} {\bibfnamefont {C.}~\bibnamefont {Bao}}, \bibinfo {author} {\bibfnamefont {M.}~\bibnamefont {Gao}},\ and\ \bibinfo {author} {\bibfnamefont {K.}~\bibnamefont {Vahala}},\ }\bibfield  {title} {\bibinfo {title} {Greater than one billion {Q} factor for on-chip microresonators},\ }\href {https://doi.org/10.1364/OL.394940} {\bibfield  {journal} {\bibinfo  {journal} {Opt. Lett.}\ }\textbf {\bibinfo {volume} {45}},\ \bibinfo {pages} {5129} (\bibinfo {year} {2020})}\BibitemShut {NoStop}%
\bibitem [{\citenamefont {Hoang}\ \emph {et~al.}(2022)\citenamefont {Hoang}, \citenamefont {Chu}, \citenamefont {Garc{\'\i}a-Vidal},\ and\ \citenamefont {Png}}]{hoang2022high}%
  \BibitemOpen
  \bibfield  {author} {\bibinfo {author} {\bibfnamefont {T.~X.}\ \bibnamefont {Hoang}}, \bibinfo {author} {\bibfnamefont {H.-S.}\ \bibnamefont {Chu}}, \bibinfo {author} {\bibfnamefont {F.~J.}\ \bibnamefont {Garc{\'\i}a-Vidal}},\ and\ \bibinfo {author} {\bibfnamefont {C.~E.}\ \bibnamefont {Png}},\ }\bibfield  {title} {\bibinfo {title} {High-performance dielectric nano-cavities for near-and mid-infrared frequency applications},\ }\href@noop {} {\bibfield  {journal} {\bibinfo  {journal} {Journal of Optics}\ }\textbf {\bibinfo {volume} {24}},\ \bibinfo {pages} {094006} (\bibinfo {year} {2022})}\BibitemShut {NoStop}%
\bibitem [{\citenamefont {Couteau}\ \emph {et~al.}(2023)\citenamefont {Couteau}, \citenamefont {Barz}, \citenamefont {Durt}, \citenamefont {Gerrits}, \citenamefont {Huwer}, \citenamefont {Prevedel}, \citenamefont {Rarity}, \citenamefont {Shields},\ and\ \citenamefont {Weihs}}]{couteau2023applications}%
  \BibitemOpen
  \bibfield  {author} {\bibinfo {author} {\bibfnamefont {C.}~\bibnamefont {Couteau}}, \bibinfo {author} {\bibfnamefont {S.}~\bibnamefont {Barz}}, \bibinfo {author} {\bibfnamefont {T.}~\bibnamefont {Durt}}, \bibinfo {author} {\bibfnamefont {T.}~\bibnamefont {Gerrits}}, \bibinfo {author} {\bibfnamefont {J.}~\bibnamefont {Huwer}}, \bibinfo {author} {\bibfnamefont {R.}~\bibnamefont {Prevedel}}, \bibinfo {author} {\bibfnamefont {J.}~\bibnamefont {Rarity}}, \bibinfo {author} {\bibfnamefont {A.}~\bibnamefont {Shields}},\ and\ \bibinfo {author} {\bibfnamefont {G.}~\bibnamefont {Weihs}},\ }\bibfield  {title} {\bibinfo {title} {Applications of single photons to quantum communication and computing},\ }\href@noop {} {\bibfield  {journal} {\bibinfo  {journal} {Nature Reviews Physics}\ }\textbf {\bibinfo {volume} {5}},\ \bibinfo {pages} {1} (\bibinfo {year} {2023})}\BibitemShut {NoStop}%
\bibitem [{\citenamefont {Kodigala}\ \emph {et~al.}(2017)\citenamefont {Kodigala}, \citenamefont {Lepetit}, \citenamefont {Gu}, \citenamefont {Bahari}, \citenamefont {Fainman},\ and\ \citenamefont {Kant{\'e}}}]{kodigala2017lasing}%
  \BibitemOpen
  \bibfield  {author} {\bibinfo {author} {\bibfnamefont {A.}~\bibnamefont {Kodigala}}, \bibinfo {author} {\bibfnamefont {T.}~\bibnamefont {Lepetit}}, \bibinfo {author} {\bibfnamefont {Q.}~\bibnamefont {Gu}}, \bibinfo {author} {\bibfnamefont {B.}~\bibnamefont {Bahari}}, \bibinfo {author} {\bibfnamefont {Y.}~\bibnamefont {Fainman}},\ and\ \bibinfo {author} {\bibfnamefont {B.}~\bibnamefont {Kant{\'e}}},\ }\bibfield  {title} {\bibinfo {title} {Lasing action from photonic bound states in continuum},\ }\href@noop {} {\bibfield  {journal} {\bibinfo  {journal} {Nature}\ }\textbf {\bibinfo {volume} {541}},\ \bibinfo {pages} {196} (\bibinfo {year} {2017})}\BibitemShut {NoStop}%
\bibitem [{\citenamefont {Ha}\ \emph {et~al.}(2018)\citenamefont {Ha}, \citenamefont {Fu}, \citenamefont {Emani}, \citenamefont {Pan}, \citenamefont {Bakker}, \citenamefont {Paniagua-Dom{\'\i}nguez},\ and\ \citenamefont {Kuznetsov}}]{ha2018directional}%
  \BibitemOpen
  \bibfield  {author} {\bibinfo {author} {\bibfnamefont {S.~T.}\ \bibnamefont {Ha}}, \bibinfo {author} {\bibfnamefont {Y.~H.}\ \bibnamefont {Fu}}, \bibinfo {author} {\bibfnamefont {N.~K.}\ \bibnamefont {Emani}}, \bibinfo {author} {\bibfnamefont {Z.}~\bibnamefont {Pan}}, \bibinfo {author} {\bibfnamefont {R.~M.}\ \bibnamefont {Bakker}}, \bibinfo {author} {\bibfnamefont {R.}~\bibnamefont {Paniagua-Dom{\'\i}nguez}},\ and\ \bibinfo {author} {\bibfnamefont {A.~I.}\ \bibnamefont {Kuznetsov}},\ }\bibfield  {title} {\bibinfo {title} {Directional lasing in resonant semiconductor nanoantenna arrays},\ }\href@noop {} {\bibfield  {journal} {\bibinfo  {journal} {Nature Nanotechnology}\ }\textbf {\bibinfo {volume} {13}},\ \bibinfo {pages} {1042} (\bibinfo {year} {2018})}\BibitemShut {NoStop}%
\bibitem [{\citenamefont {Liu}\ \emph {et~al.}(2019)\citenamefont {Liu}, \citenamefont {Xu}, \citenamefont {Lin}, \citenamefont {Xiang}, \citenamefont {Feng}, \citenamefont {Cao}, \citenamefont {Li}, \citenamefont {Lan},\ and\ \citenamefont {Liu}}]{liu2019high}%
  \BibitemOpen
  \bibfield  {author} {\bibinfo {author} {\bibfnamefont {Z.}~\bibnamefont {Liu}}, \bibinfo {author} {\bibfnamefont {Y.}~\bibnamefont {Xu}}, \bibinfo {author} {\bibfnamefont {Y.}~\bibnamefont {Lin}}, \bibinfo {author} {\bibfnamefont {J.}~\bibnamefont {Xiang}}, \bibinfo {author} {\bibfnamefont {T.}~\bibnamefont {Feng}}, \bibinfo {author} {\bibfnamefont {Q.}~\bibnamefont {Cao}}, \bibinfo {author} {\bibfnamefont {J.}~\bibnamefont {Li}}, \bibinfo {author} {\bibfnamefont {S.}~\bibnamefont {Lan}},\ and\ \bibinfo {author} {\bibfnamefont {J.}~\bibnamefont {Liu}},\ }\bibfield  {title} {\bibinfo {title} {High-{Q} quasibound states in the continuum for nonlinear metasurfaces},\ }\href@noop {} {\bibfield  {journal} {\bibinfo  {journal} {Physical Review Letters}\ }\textbf {\bibinfo {volume} {123}},\ \bibinfo {pages} {253901} (\bibinfo {year} {2019})}\BibitemShut {NoStop}%
\bibitem [{\citenamefont {Johnson}\ \emph {et~al.}(1999)\citenamefont {Johnson}, \citenamefont {Fan}, \citenamefont {Villeneuve}, \citenamefont {Joannopoulos},\ and\ \citenamefont {Kolodziejski}}]{johnson1999guided}%
  \BibitemOpen
  \bibfield  {author} {\bibinfo {author} {\bibfnamefont {S.~G.}\ \bibnamefont {Johnson}}, \bibinfo {author} {\bibfnamefont {S.}~\bibnamefont {Fan}}, \bibinfo {author} {\bibfnamefont {P.~R.}\ \bibnamefont {Villeneuve}}, \bibinfo {author} {\bibfnamefont {J.~D.}\ \bibnamefont {Joannopoulos}},\ and\ \bibinfo {author} {\bibfnamefont {L.}~\bibnamefont {Kolodziejski}},\ }\bibfield  {title} {\bibinfo {title} {Guided modes in photonic crystal slabs},\ }\href@noop {} {\bibfield  {journal} {\bibinfo  {journal} {Physical Review B}\ }\textbf {\bibinfo {volume} {60}},\ \bibinfo {pages} {5751} (\bibinfo {year} {1999})}\BibitemShut {NoStop}%
\bibitem [{\citenamefont {Fan}\ and\ \citenamefont {Joannopoulos}(2002)}]{fan2002analysis}%
  \BibitemOpen
  \bibfield  {author} {\bibinfo {author} {\bibfnamefont {S.}~\bibnamefont {Fan}}\ and\ \bibinfo {author} {\bibfnamefont {J.~D.}\ \bibnamefont {Joannopoulos}},\ }\bibfield  {title} {\bibinfo {title} {Analysis of guided resonances in photonic crystal slabs},\ }\href@noop {} {\bibfield  {journal} {\bibinfo  {journal} {Physical Review B}\ }\textbf {\bibinfo {volume} {65}},\ \bibinfo {pages} {235112} (\bibinfo {year} {2002})}\BibitemShut {NoStop}%
\bibitem [{\citenamefont {Vaishnav}\ \emph {et~al.}(2007)\citenamefont {Vaishnav}, \citenamefont {Walls}, \citenamefont {Apratim},\ and\ \citenamefont {Heller}}]{vaishnav2007matter}%
  \BibitemOpen
  \bibfield  {author} {\bibinfo {author} {\bibfnamefont {J.}~\bibnamefont {Vaishnav}}, \bibinfo {author} {\bibfnamefont {J.}~\bibnamefont {Walls}}, \bibinfo {author} {\bibfnamefont {M.}~\bibnamefont {Apratim}},\ and\ \bibinfo {author} {\bibfnamefont {E.}~\bibnamefont {Heller}},\ }\bibfield  {title} {\bibinfo {title} {Matter-wave scattering and guiding by atomic arrays},\ }\href@noop {} {\bibfield  {journal} {\bibinfo  {journal} {Physical Review A}\ }\textbf {\bibinfo {volume} {76}},\ \bibinfo {pages} {013620} (\bibinfo {year} {2007})}\BibitemShut {NoStop}%
\bibitem [{\citenamefont {Hsu}\ \emph {et~al.}(2013)\citenamefont {Hsu}, \citenamefont {Zhen}, \citenamefont {Lee}, \citenamefont {Chua}, \citenamefont {Johnson}, \citenamefont {Joannopoulos},\ and\ \citenamefont {Solja{\v{c}}i{\'c}}}]{hsu2013observation}%
  \BibitemOpen
  \bibfield  {author} {\bibinfo {author} {\bibfnamefont {C.~W.}\ \bibnamefont {Hsu}}, \bibinfo {author} {\bibfnamefont {B.}~\bibnamefont {Zhen}}, \bibinfo {author} {\bibfnamefont {J.}~\bibnamefont {Lee}}, \bibinfo {author} {\bibfnamefont {S.-L.}\ \bibnamefont {Chua}}, \bibinfo {author} {\bibfnamefont {S.~G.}\ \bibnamefont {Johnson}}, \bibinfo {author} {\bibfnamefont {J.~D.}\ \bibnamefont {Joannopoulos}},\ and\ \bibinfo {author} {\bibfnamefont {M.}~\bibnamefont {Solja{\v{c}}i{\'c}}},\ }\bibfield  {title} {\bibinfo {title} {Observation of trapped light within the radiation continuum},\ }\href@noop {} {\bibfield  {journal} {\bibinfo  {journal} {Nature}\ }\textbf {\bibinfo {volume} {499}},\ \bibinfo {pages} {188} (\bibinfo {year} {2013})}\BibitemShut {NoStop}%
\bibitem [{\citenamefont {Zhen}\ \emph {et~al.}(2014)\citenamefont {Zhen}, \citenamefont {Hsu}, \citenamefont {Lu}, \citenamefont {Stone},\ and\ \citenamefont {Solja{\v{c}}i{\'c}}}]{zhen2014topological}%
  \BibitemOpen
  \bibfield  {author} {\bibinfo {author} {\bibfnamefont {B.}~\bibnamefont {Zhen}}, \bibinfo {author} {\bibfnamefont {C.~W.}\ \bibnamefont {Hsu}}, \bibinfo {author} {\bibfnamefont {L.}~\bibnamefont {Lu}}, \bibinfo {author} {\bibfnamefont {A.~D.}\ \bibnamefont {Stone}},\ and\ \bibinfo {author} {\bibfnamefont {M.}~\bibnamefont {Solja{\v{c}}i{\'c}}},\ }\bibfield  {title} {\bibinfo {title} {Topological nature of optical bound states in the continuum},\ }\href@noop {} {\bibfield  {journal} {\bibinfo  {journal} {Physical Review Letters}\ }\textbf {\bibinfo {volume} {113}},\ \bibinfo {pages} {257401} (\bibinfo {year} {2014})}\BibitemShut {NoStop}%
\bibitem [{\citenamefont {Bulgakov}\ and\ \citenamefont {Maksimov}(2017)}]{bulgakov2017topological}%
  \BibitemOpen
  \bibfield  {author} {\bibinfo {author} {\bibfnamefont {E.~N.}\ \bibnamefont {Bulgakov}}\ and\ \bibinfo {author} {\bibfnamefont {D.~N.}\ \bibnamefont {Maksimov}},\ }\bibfield  {title} {\bibinfo {title} {Topological bound states in the continuum in arrays of dielectric spheres},\ }\href@noop {} {\bibfield  {journal} {\bibinfo  {journal} {Physical Review Letters}\ }\textbf {\bibinfo {volume} {118}},\ \bibinfo {pages} {267401} (\bibinfo {year} {2017})}\BibitemShut {NoStop}%
\bibitem [{\citenamefont {Lagendijk}\ and\ \citenamefont {Van~Tiggelen}(1996)}]{lagendijk1996resonant}%
  \BibitemOpen
  \bibfield  {author} {\bibinfo {author} {\bibfnamefont {A.}~\bibnamefont {Lagendijk}}\ and\ \bibinfo {author} {\bibfnamefont {B.~A.}\ \bibnamefont {Van~Tiggelen}},\ }\bibfield  {title} {\bibinfo {title} {Resonant multiple scattering of light},\ }\href@noop {} {\bibfield  {journal} {\bibinfo  {journal} {Physics Reports}\ }\textbf {\bibinfo {volume} {270}},\ \bibinfo {pages} {143} (\bibinfo {year} {1996})}\BibitemShut {NoStop}%
\bibitem [{\citenamefont {Devaney}\ and\ \citenamefont {Wolf}(1974)}]{devaney1974multipole}%
  \BibitemOpen
  \bibfield  {author} {\bibinfo {author} {\bibfnamefont {A.}~\bibnamefont {Devaney}}\ and\ \bibinfo {author} {\bibfnamefont {E.}~\bibnamefont {Wolf}},\ }\bibfield  {title} {\bibinfo {title} {Multipole expansions and plane wave representations of the electromagnetic field},\ }\href@noop {} {\bibfield  {journal} {\bibinfo  {journal} {Journal of Mathematical Physics}\ }\textbf {\bibinfo {volume} {15}},\ \bibinfo {pages} {234} (\bibinfo {year} {1974})}\BibitemShut {NoStop}%
\bibitem [{\citenamefont {Hoang}\ \emph {et~al.}(2014)\citenamefont {Hoang}, \citenamefont {Chen},\ and\ \citenamefont {Sheppard}}]{hoang2014multipole}%
  \BibitemOpen
  \bibfield  {author} {\bibinfo {author} {\bibfnamefont {T.~X.}\ \bibnamefont {Hoang}}, \bibinfo {author} {\bibfnamefont {X.}~\bibnamefont {Chen}},\ and\ \bibinfo {author} {\bibfnamefont {C.~J.}\ \bibnamefont {Sheppard}},\ }\bibfield  {title} {\bibinfo {title} {Multipole and plane wave expansions of diverging and converging fields},\ }\href@noop {} {\bibfield  {journal} {\bibinfo  {journal} {Optics Express}\ }\textbf {\bibinfo {volume} {22}},\ \bibinfo {pages} {8949} (\bibinfo {year} {2014})}\BibitemShut {NoStop}%
\bibitem [{\citenamefont {Hoang}\ \emph {et~al.}(2017)\citenamefont {Hoang}, \citenamefont {Nagelberg}, \citenamefont {Kolle},\ and\ \citenamefont {Barbastathis}}]{hoang2017fano}%
  \BibitemOpen
  \bibfield  {author} {\bibinfo {author} {\bibfnamefont {T.~X.}\ \bibnamefont {Hoang}}, \bibinfo {author} {\bibfnamefont {S.~N.}\ \bibnamefont {Nagelberg}}, \bibinfo {author} {\bibfnamefont {M.}~\bibnamefont {Kolle}},\ and\ \bibinfo {author} {\bibfnamefont {G.}~\bibnamefont {Barbastathis}},\ }\bibfield  {title} {\bibinfo {title} {Fano resonances from coupled whispering--gallery modes in photonic molecules},\ }\href@noop {} {\bibfield  {journal} {\bibinfo  {journal} {Optics Express}\ }\textbf {\bibinfo {volume} {25}},\ \bibinfo {pages} {13125} (\bibinfo {year} {2017})}\BibitemShut {NoStop}%
\bibitem [{\citenamefont {Sun}\ \emph {et~al.}(2024)\citenamefont {Sun}, \citenamefont {Wang},\ and\ \citenamefont {Han}}]{sun2024high}%
  \BibitemOpen
  \bibfield  {author} {\bibinfo {author} {\bibfnamefont {K.}~\bibnamefont {Sun}}, \bibinfo {author} {\bibfnamefont {W.}~\bibnamefont {Wang}},\ and\ \bibinfo {author} {\bibfnamefont {Z.}~\bibnamefont {Han}},\ }\bibfield  {title} {\bibinfo {title} {High-q resonances in periodic photonic structures},\ }\href@noop {} {\bibfield  {journal} {\bibinfo  {journal} {Physical Review B}\ }\textbf {\bibinfo {volume} {109}},\ \bibinfo {pages} {085426} (\bibinfo {year} {2024})}\BibitemShut {NoStop}%
\bibitem [{\citenamefont {Sidorenko}\ \emph {et~al.}(2021)\citenamefont {Sidorenko}, \citenamefont {Sergaeva}, \citenamefont {Sadrieva}, \citenamefont {Roques-Carmes}, \citenamefont {Muraev}, \citenamefont {Maksimov},\ and\ \citenamefont {Bogdanov}}]{sidorenko2021observation}%
  \BibitemOpen
  \bibfield  {author} {\bibinfo {author} {\bibfnamefont {M.}~\bibnamefont {Sidorenko}}, \bibinfo {author} {\bibfnamefont {O.}~\bibnamefont {Sergaeva}}, \bibinfo {author} {\bibfnamefont {Z.}~\bibnamefont {Sadrieva}}, \bibinfo {author} {\bibfnamefont {C.}~\bibnamefont {Roques-Carmes}}, \bibinfo {author} {\bibfnamefont {P.}~\bibnamefont {Muraev}}, \bibinfo {author} {\bibfnamefont {D.}~\bibnamefont {Maksimov}},\ and\ \bibinfo {author} {\bibfnamefont {A.}~\bibnamefont {Bogdanov}},\ }\bibfield  {title} {\bibinfo {title} {Observation of an accidental bound state in the continuum in a chain of dielectric disks},\ }\href@noop {} {\bibfield  {journal} {\bibinfo  {journal} {Physical Review Applied}\ }\textbf {\bibinfo {volume} {15}},\ \bibinfo {pages} {034041} (\bibinfo {year} {2021})}\BibitemShut {NoStop}%
\bibitem [{\citenamefont {Hartmann}\ \emph {et~al.}(2006)\citenamefont {Hartmann}, \citenamefont {Brandao},\ and\ \citenamefont {Plenio}}]{hartmann2006strongly}%
  \BibitemOpen
  \bibfield  {author} {\bibinfo {author} {\bibfnamefont {M.~J.}\ \bibnamefont {Hartmann}}, \bibinfo {author} {\bibfnamefont {F.~G.}\ \bibnamefont {Brandao}},\ and\ \bibinfo {author} {\bibfnamefont {M.~B.}\ \bibnamefont {Plenio}},\ }\bibfield  {title} {\bibinfo {title} {Strongly interacting polaritons in coupled arrays of cavities},\ }\href@noop {} {\bibfield  {journal} {\bibinfo  {journal} {Nature Physics}\ }\textbf {\bibinfo {volume} {2}},\ \bibinfo {pages} {849} (\bibinfo {year} {2006})}\BibitemShut {NoStop}%
\bibitem [{\citenamefont {Reiserer}\ and\ \citenamefont {Rempe}(2015)}]{reiserer2015cavity}%
  \BibitemOpen
  \bibfield  {author} {\bibinfo {author} {\bibfnamefont {A.}~\bibnamefont {Reiserer}}\ and\ \bibinfo {author} {\bibfnamefont {G.}~\bibnamefont {Rempe}},\ }\bibfield  {title} {\bibinfo {title} {Cavity-based quantum networks with single atoms and optical photons},\ }\href@noop {} {\bibfield  {journal} {\bibinfo  {journal} {Reviews of Modern Physics}\ }\textbf {\bibinfo {volume} {87}},\ \bibinfo {pages} {1379} (\bibinfo {year} {2015})}\BibitemShut {NoStop}%
\bibitem [{\citenamefont {Chang}\ \emph {et~al.}(2018)\citenamefont {Chang}, \citenamefont {Douglas}, \citenamefont {Gonz{\'a}lez-Tudela}, \citenamefont {Hung},\ and\ \citenamefont {Kimble}}]{chang2018colloquium}%
  \BibitemOpen
  \bibfield  {author} {\bibinfo {author} {\bibfnamefont {D.}~\bibnamefont {Chang}}, \bibinfo {author} {\bibfnamefont {J.}~\bibnamefont {Douglas}}, \bibinfo {author} {\bibfnamefont {A.}~\bibnamefont {Gonz{\'a}lez-Tudela}}, \bibinfo {author} {\bibfnamefont {C.-L.}\ \bibnamefont {Hung}},\ and\ \bibinfo {author} {\bibfnamefont {H.}~\bibnamefont {Kimble}},\ }\bibfield  {title} {\bibinfo {title} {Colloquium: Quantum matter built from nanoscopic lattices of atoms and photons},\ }\href@noop {} {\bibfield  {journal} {\bibinfo  {journal} {Reviews of Modern Physics}\ }\textbf {\bibinfo {volume} {90}},\ \bibinfo {pages} {031002} (\bibinfo {year} {2018})}\BibitemShut {NoStop}%
\bibitem [{\citenamefont {Sadrieva}\ \emph {et~al.}(2019)\citenamefont {Sadrieva}, \citenamefont {Frizyuk}, \citenamefont {Petrov}, \citenamefont {Kivshar},\ and\ \citenamefont {Bogdanov}}]{sadrieva2019multipolar}%
  \BibitemOpen
  \bibfield  {author} {\bibinfo {author} {\bibfnamefont {Z.}~\bibnamefont {Sadrieva}}, \bibinfo {author} {\bibfnamefont {K.}~\bibnamefont {Frizyuk}}, \bibinfo {author} {\bibfnamefont {M.}~\bibnamefont {Petrov}}, \bibinfo {author} {\bibfnamefont {Y.}~\bibnamefont {Kivshar}},\ and\ \bibinfo {author} {\bibfnamefont {A.}~\bibnamefont {Bogdanov}},\ }\bibfield  {title} {\bibinfo {title} {Multipolar origin of bound states in the continuum},\ }\href@noop {} {\bibfield  {journal} {\bibinfo  {journal} {Physical Review B}\ }\textbf {\bibinfo {volume} {100}},\ \bibinfo {pages} {115303} (\bibinfo {year} {2019})}\BibitemShut {NoStop}%
\bibitem [{\citenamefont {Ustimenko}\ \emph {et~al.}(2024)\citenamefont {Ustimenko}, \citenamefont {Rockstuhl},\ and\ \citenamefont {Evlyukhin}}]{ustimenko2024resonances}%
  \BibitemOpen
  \bibfield  {author} {\bibinfo {author} {\bibfnamefont {N.}~\bibnamefont {Ustimenko}}, \bibinfo {author} {\bibfnamefont {C.}~\bibnamefont {Rockstuhl}},\ and\ \bibinfo {author} {\bibfnamefont {A.~B.}\ \bibnamefont {Evlyukhin}},\ }\bibfield  {title} {\bibinfo {title} {Resonances in finite-size all-dielectric metasurfaces for light trapping and propagation control},\ }\href@noop {} {\bibfield  {journal} {\bibinfo  {journal} {Physical Review B}\ }\textbf {\bibinfo {volume} {109}},\ \bibinfo {pages} {115436} (\bibinfo {year} {2024})}\BibitemShut {NoStop}%
\bibitem [{\citenamefont {Mikhailovskii}\ \emph {et~al.}(2024)\citenamefont {Mikhailovskii}, \citenamefont {Poleva}, \citenamefont {Solodovchenko}, \citenamefont {Sidorenko}, \citenamefont {Sadrieva}, \citenamefont {Petrov}, \citenamefont {Bogdanov},\ and\ \citenamefont {Savelev}}]{mikhailovskii2024engineering}%
  \BibitemOpen
  \bibfield  {author} {\bibinfo {author} {\bibfnamefont {M.}~\bibnamefont {Mikhailovskii}}, \bibinfo {author} {\bibfnamefont {M.}~\bibnamefont {Poleva}}, \bibinfo {author} {\bibfnamefont {N.}~\bibnamefont {Solodovchenko}}, \bibinfo {author} {\bibfnamefont {M.}~\bibnamefont {Sidorenko}}, \bibinfo {author} {\bibfnamefont {Z.}~\bibnamefont {Sadrieva}}, \bibinfo {author} {\bibfnamefont {M.}~\bibnamefont {Petrov}}, \bibinfo {author} {\bibfnamefont {A.}~\bibnamefont {Bogdanov}},\ and\ \bibinfo {author} {\bibfnamefont {R.}~\bibnamefont {Savelev}},\ }\bibfield  {title} {\bibinfo {title} {Engineering of high‑q states via collective mode coupling in chains of mie resonators},\ }\href@noop {} {\bibfield  {journal} {\bibinfo  {journal} {ACS Photonics}\ }\textbf {\bibinfo {volume} {11}},\ \bibinfo {pages} {1657} (\bibinfo {year} {2024})}\BibitemShut {NoStop}%
\bibitem [{\citenamefont {Overvig}\ \emph {et~al.}(2020)\citenamefont {Overvig}, \citenamefont {Malek}, \citenamefont {Carter}, \citenamefont {Shrestha},\ and\ \citenamefont {Yu}}]{overvig2020selection}%
  \BibitemOpen
  \bibfield  {author} {\bibinfo {author} {\bibfnamefont {A.~C.}\ \bibnamefont {Overvig}}, \bibinfo {author} {\bibfnamefont {S.~C.}\ \bibnamefont {Malek}}, \bibinfo {author} {\bibfnamefont {M.~J.}\ \bibnamefont {Carter}}, \bibinfo {author} {\bibfnamefont {S.}~\bibnamefont {Shrestha}},\ and\ \bibinfo {author} {\bibfnamefont {N.}~\bibnamefont {Yu}},\ }\bibfield  {title} {\bibinfo {title} {Selection rules for quasibound states in the continuum},\ }\href@noop {} {\bibfield  {journal} {\bibinfo  {journal} {Physical Review B}\ }\textbf {\bibinfo {volume} {102}},\ \bibinfo {pages} {035434} (\bibinfo {year} {2020})}\BibitemShut {NoStop}%
\bibitem [{\citenamefont {Geints}(2023{\natexlab{a}})}]{geints2023phase}%
  \BibitemOpen
  \bibfield  {author} {\bibinfo {author} {\bibfnamefont {Y.~E.}\ \bibnamefont {Geints}},\ }\bibfield  {title} {\bibinfo {title} {Phase-controlled supermodes in symmetric photonic molecules},\ }\href@noop {} {\bibfield  {journal} {\bibinfo  {journal} {Journal of Quantitative Spectroscopy and Radiative Transfer}\ }\textbf {\bibinfo {volume} {302}},\ \bibinfo {pages} {108524} (\bibinfo {year} {2023}{\natexlab{a}})}\BibitemShut {NoStop}%
\bibitem [{\citenamefont {Geints}(2023{\natexlab{b}})}]{geints2023manipulating}%
  \BibitemOpen
  \bibfield  {author} {\bibinfo {author} {\bibfnamefont {Y.~E.}\ \bibnamefont {Geints}},\ }\bibfield  {title} {\bibinfo {title} {Manipulating the supermodes in photonic molecules: prospects for all-optical switching and sensing},\ }\href@noop {} {\bibfield  {journal} {\bibinfo  {journal} {JOSA B}\ }\textbf {\bibinfo {volume} {40}},\ \bibinfo {pages} {1875} (\bibinfo {year} {2023}{\natexlab{b}})}\BibitemShut {NoStop}%
\bibitem [{\citenamefont {Feshbach}(1958)}]{feshbach1958unified}%
  \BibitemOpen
  \bibfield  {author} {\bibinfo {author} {\bibfnamefont {H.}~\bibnamefont {Feshbach}},\ }\bibfield  {title} {\bibinfo {title} {Unified theory of nuclear reactions},\ }\href@noop {} {\bibfield  {journal} {\bibinfo  {journal} {Annals of Physics}\ }\textbf {\bibinfo {volume} {5}},\ \bibinfo {pages} {357} (\bibinfo {year} {1958})}\BibitemShut {NoStop}%
\bibitem [{\citenamefont {Bulgakov}\ and\ \citenamefont {Sadreev}(2019)}]{bulgakov2019high}%
  \BibitemOpen
  \bibfield  {author} {\bibinfo {author} {\bibfnamefont {E.~N.}\ \bibnamefont {Bulgakov}}\ and\ \bibinfo {author} {\bibfnamefont {A.~F.}\ \bibnamefont {Sadreev}},\ }\bibfield  {title} {\bibinfo {title} {High-{Q} resonant modes in a finite array of dielectric particles},\ }\href@noop {} {\bibfield  {journal} {\bibinfo  {journal} {Physical Review A}\ }\textbf {\bibinfo {volume} {99}},\ \bibinfo {pages} {033851} (\bibinfo {year} {2019})}\BibitemShut {NoStop}%
\bibitem [{\citenamefont {Hoang}\ \emph {et~al.}(2024)\citenamefont {Hoang}, \citenamefont {Leykam},\ and\ \citenamefont {Kivshar}}]{hoang2024photonic}%
  \BibitemOpen
  \bibfield  {author} {\bibinfo {author} {\bibfnamefont {T.~X.}\ \bibnamefont {Hoang}}, \bibinfo {author} {\bibfnamefont {D.}~\bibnamefont {Leykam}},\ and\ \bibinfo {author} {\bibfnamefont {Y.}~\bibnamefont {Kivshar}},\ }\bibfield  {title} {\bibinfo {title} {Photonic flatband resonances in multiple light scattering},\ }\href@noop {} {\bibfield  {journal} {\bibinfo  {journal} {Physical Review Letters}\ }\textbf {\bibinfo {volume} {132}},\ \bibinfo {pages} {043803} (\bibinfo {year} {2024})}\BibitemShut {NoStop}%
\bibitem [{\citenamefont {Hoang}\ \emph {et~al.}(2012)\citenamefont {Hoang}, \citenamefont {Chen},\ and\ \citenamefont {Sheppard}}]{hoang2012multipole}%
  \BibitemOpen
  \bibfield  {author} {\bibinfo {author} {\bibfnamefont {T.~X.}\ \bibnamefont {Hoang}}, \bibinfo {author} {\bibfnamefont {X.}~\bibnamefont {Chen}},\ and\ \bibinfo {author} {\bibfnamefont {C.~J.}\ \bibnamefont {Sheppard}},\ }\bibfield  {title} {\bibinfo {title} {Multipole theory for tight focusing of polarized light, including radially polarized and other special cases},\ }\href@noop {} {\bibfield  {journal} {\bibinfo  {journal} {JOSA A}\ }\textbf {\bibinfo {volume} {29}},\ \bibinfo {pages} {32} (\bibinfo {year} {2012})}\BibitemShut {NoStop}%
\bibitem [{\citenamefont {Bliokh}\ \emph {et~al.}(2023)\citenamefont {Bliokh}, \citenamefont {Karimi}, \citenamefont {Padgett}, \citenamefont {Alonso}, \citenamefont {Dennis}, \citenamefont {Dudley}, \citenamefont {Forbes}, \citenamefont {Zahedpour}, \citenamefont {Hancock}, \citenamefont {Milchberg} \emph {et~al.}}]{bliokh2023roadmap}%
  \BibitemOpen
  \bibfield  {author} {\bibinfo {author} {\bibfnamefont {K.~Y.}\ \bibnamefont {Bliokh}}, \bibinfo {author} {\bibfnamefont {E.}~\bibnamefont {Karimi}}, \bibinfo {author} {\bibfnamefont {M.~J.}\ \bibnamefont {Padgett}}, \bibinfo {author} {\bibfnamefont {M.~A.}\ \bibnamefont {Alonso}}, \bibinfo {author} {\bibfnamefont {M.~R.}\ \bibnamefont {Dennis}}, \bibinfo {author} {\bibfnamefont {A.}~\bibnamefont {Dudley}}, \bibinfo {author} {\bibfnamefont {A.}~\bibnamefont {Forbes}}, \bibinfo {author} {\bibfnamefont {S.}~\bibnamefont {Zahedpour}}, \bibinfo {author} {\bibfnamefont {S.~W.}\ \bibnamefont {Hancock}}, \bibinfo {author} {\bibfnamefont {H.~M.}\ \bibnamefont {Milchberg}}, \emph {et~al.},\ }\bibfield  {title} {\bibinfo {title} {Roadmap on structured waves},\ }\href@noop {} {\bibfield  {journal} {\bibinfo  {journal} {Journal of Optics}\ }\textbf {\bibinfo {volume} {25}},\ \bibinfo {pages} {103001} (\bibinfo {year} {2023})}\BibitemShut {NoStop}%
\bibitem [{\citenamefont {Zhang}\ and\ \citenamefont {M{\o}lmer}(2020)}]{zhang2020subradiant}%
  \BibitemOpen
  \bibfield  {author} {\bibinfo {author} {\bibfnamefont {Y.-X.}\ \bibnamefont {Zhang}}\ and\ \bibinfo {author} {\bibfnamefont {K.}~\bibnamefont {M{\o}lmer}},\ }\bibfield  {title} {\bibinfo {title} {Subradiant emission from regular atomic arrays: Universal scaling of decay rates from the generalized {B}loch theorem},\ }\href@noop {} {\bibfield  {journal} {\bibinfo  {journal} {Physical Review Letters}\ }\textbf {\bibinfo {volume} {125}},\ \bibinfo {pages} {253601} (\bibinfo {year} {2020})}\BibitemShut {NoStop}%
\bibitem [{\citenamefont {Volkov}\ \emph {et~al.}(2024)\citenamefont {Volkov}, \citenamefont {Ustimenko}, \citenamefont {Kornovan}, \citenamefont {Sheremet}, \citenamefont {Savelev},\ and\ \citenamefont {Petrov}}]{volkov2024strongly}%
  \BibitemOpen
  \bibfield  {author} {\bibinfo {author} {\bibfnamefont {I.~A.}\ \bibnamefont {Volkov}}, \bibinfo {author} {\bibfnamefont {N.~A.}\ \bibnamefont {Ustimenko}}, \bibinfo {author} {\bibfnamefont {D.~F.}\ \bibnamefont {Kornovan}}, \bibinfo {author} {\bibfnamefont {A.~S.}\ \bibnamefont {Sheremet}}, \bibinfo {author} {\bibfnamefont {R.~S.}\ \bibnamefont {Savelev}},\ and\ \bibinfo {author} {\bibfnamefont {M.~I.}\ \bibnamefont {Petrov}},\ }\bibfield  {title} {\bibinfo {title} {Strongly subradiant states in planar atomic arrays},\ }\href@noop {} {\bibfield  {journal} {\bibinfo  {journal} {Nanophotonics}\ }\textbf {\bibinfo {volume} {13}},\ \bibinfo {pages} {289} (\bibinfo {year} {2024})}\BibitemShut {NoStop}%
\bibitem [{\citenamefont {Asselie}\ \emph {et~al.}(2022)\citenamefont {Asselie}, \citenamefont {Cipris},\ and\ \citenamefont {Guerin}}]{asselie2022optical}%
  \BibitemOpen
  \bibfield  {author} {\bibinfo {author} {\bibfnamefont {S.}~\bibnamefont {Asselie}}, \bibinfo {author} {\bibfnamefont {A.}~\bibnamefont {Cipris}},\ and\ \bibinfo {author} {\bibfnamefont {W.}~\bibnamefont {Guerin}},\ }\bibfield  {title} {\bibinfo {title} {Optical interpretation of linear-optics superradiance and subradiance},\ }\href@noop {} {\bibfield  {journal} {\bibinfo  {journal} {Physical Review A}\ }\textbf {\bibinfo {volume} {106}},\ \bibinfo {pages} {063712} (\bibinfo {year} {2022})}\BibitemShut {NoStop}%
\bibitem [{\citenamefont {Stillinger}\ and\ \citenamefont {Herrick}(1975)}]{stillinger1975bound}%
  \BibitemOpen
  \bibfield  {author} {\bibinfo {author} {\bibfnamefont {F.~H.}\ \bibnamefont {Stillinger}}\ and\ \bibinfo {author} {\bibfnamefont {D.~R.}\ \bibnamefont {Herrick}},\ }\bibfield  {title} {\bibinfo {title} {Bound states in the continuum},\ }\href@noop {} {\bibfield  {journal} {\bibinfo  {journal} {Physical Review A}\ }\textbf {\bibinfo {volume} {11}},\ \bibinfo {pages} {446} (\bibinfo {year} {1975})}\BibitemShut {NoStop}%
\bibitem [{\citenamefont {Capasso}\ \emph {et~al.}(1992)\citenamefont {Capasso}, \citenamefont {Sirtori}, \citenamefont {Faist}, \citenamefont {Sivco}, \citenamefont {Chu},\ and\ \citenamefont {Cho}}]{capasso1992observation}%
  \BibitemOpen
  \bibfield  {author} {\bibinfo {author} {\bibfnamefont {F.}~\bibnamefont {Capasso}}, \bibinfo {author} {\bibfnamefont {C.}~\bibnamefont {Sirtori}}, \bibinfo {author} {\bibfnamefont {J.}~\bibnamefont {Faist}}, \bibinfo {author} {\bibfnamefont {D.~L.}\ \bibnamefont {Sivco}}, \bibinfo {author} {\bibfnamefont {S.-N.~G.}\ \bibnamefont {Chu}},\ and\ \bibinfo {author} {\bibfnamefont {A.~Y.}\ \bibnamefont {Cho}},\ }\bibfield  {title} {\bibinfo {title} {Observation of an electronic bound state above a potential well},\ }\href@noop {} {\bibfield  {journal} {\bibinfo  {journal} {Nature}\ }\textbf {\bibinfo {volume} {358}},\ \bibinfo {pages} {565} (\bibinfo {year} {1992})}\BibitemShut {NoStop}%
\bibitem [{\citenamefont {Stillinger}(1976)}]{stillinger1976potentials}%
  \BibitemOpen
  \bibfield  {author} {\bibinfo {author} {\bibfnamefont {F.}~\bibnamefont {Stillinger}},\ }\bibfield  {title} {\bibinfo {title} {Potentials supporting positive-energy eigenstates and their application to semiconductor heterostructures},\ }\href@noop {} {\bibfield  {journal} {\bibinfo  {journal} {Physica B+C}\ }\textbf {\bibinfo {volume} {85}},\ \bibinfo {pages} {270} (\bibinfo {year} {1976})}\BibitemShut {NoStop}%
\bibitem [{\citenamefont {Fonda}(1963)}]{fonda1963bound}%
  \BibitemOpen
  \bibfield  {author} {\bibinfo {author} {\bibfnamefont {L.}~\bibnamefont {Fonda}},\ }\bibfield  {title} {\bibinfo {title} {Bound states embedded in the continuum and the formal theory of scattering},\ }\href@noop {} {\bibfield  {journal} {\bibinfo  {journal} {Annals of Physics}\ }\textbf {\bibinfo {volume} {22}},\ \bibinfo {pages} {123} (\bibinfo {year} {1963})}\BibitemShut {NoStop}%
\bibitem [{\citenamefont {Friedrich}\ and\ \citenamefont {Wintgen}(1985)}]{friedrich1985interfering}%
  \BibitemOpen
  \bibfield  {author} {\bibinfo {author} {\bibfnamefont {H.}~\bibnamefont {Friedrich}}\ and\ \bibinfo {author} {\bibfnamefont {D.}~\bibnamefont {Wintgen}},\ }\bibfield  {title} {\bibinfo {title} {Interfering resonances and bound states in the continuum},\ }\href@noop {} {\bibfield  {journal} {\bibinfo  {journal} {Physical Review A}\ }\textbf {\bibinfo {volume} {32}},\ \bibinfo {pages} {3231} (\bibinfo {year} {1985})}\BibitemShut {NoStop}%
\bibitem [{\citenamefont {Marinica}\ \emph {et~al.}(2008)\citenamefont {Marinica}, \citenamefont {Borisov},\ and\ \citenamefont {Shabanov}}]{marinica2008bound}%
  \BibitemOpen
  \bibfield  {author} {\bibinfo {author} {\bibfnamefont {D.}~\bibnamefont {Marinica}}, \bibinfo {author} {\bibfnamefont {A.}~\bibnamefont {Borisov}},\ and\ \bibinfo {author} {\bibfnamefont {S.}~\bibnamefont {Shabanov}},\ }\bibfield  {title} {\bibinfo {title} {Bound states in the continuum in photonics},\ }\href@noop {} {\bibfield  {journal} {\bibinfo  {journal} {Physical Review Letters}\ }\textbf {\bibinfo {volume} {100}},\ \bibinfo {pages} {183902} (\bibinfo {year} {2008})}\BibitemShut {NoStop}%
\bibitem [{\citenamefont {Dong}\ \emph {et~al.}(2022)\citenamefont {Dong}, \citenamefont {Mahfoud}, \citenamefont {Paniagua-Dom{\'\i}nguez}, \citenamefont {Wang}, \citenamefont {Fern{\'a}ndez-Dom{\'\i}nguez}, \citenamefont {Gorelik}, \citenamefont {Ha}, \citenamefont {Tjiptoharsono}, \citenamefont {Kuznetsov}, \citenamefont {Bosman} \emph {et~al.}}]{dong2022nanoscale}%
  \BibitemOpen
  \bibfield  {author} {\bibinfo {author} {\bibfnamefont {Z.}~\bibnamefont {Dong}}, \bibinfo {author} {\bibfnamefont {Z.}~\bibnamefont {Mahfoud}}, \bibinfo {author} {\bibfnamefont {R.}~\bibnamefont {Paniagua-Dom{\'\i}nguez}}, \bibinfo {author} {\bibfnamefont {H.}~\bibnamefont {Wang}}, \bibinfo {author} {\bibfnamefont {A.~I.}\ \bibnamefont {Fern{\'a}ndez-Dom{\'\i}nguez}}, \bibinfo {author} {\bibfnamefont {S.}~\bibnamefont {Gorelik}}, \bibinfo {author} {\bibfnamefont {S.~T.}\ \bibnamefont {Ha}}, \bibinfo {author} {\bibfnamefont {F.}~\bibnamefont {Tjiptoharsono}}, \bibinfo {author} {\bibfnamefont {A.~I.}\ \bibnamefont {Kuznetsov}}, \bibinfo {author} {\bibfnamefont {M.}~\bibnamefont {Bosman}}, \emph {et~al.},\ }\bibfield  {title} {\bibinfo {title} {Nanoscale mapping of optically inaccessible bound-states-in-the-continuum},\ }\href@noop {} {\bibfield  {journal} {\bibinfo  {journal} {Light: Science \& Applications}\ }\textbf {\bibinfo {volume} {11}},\ \bibinfo {pages} {20} (\bibinfo {year} {2022})}\BibitemShut
  {NoStop}%
\bibitem [{\citenamefont {Xu}\ \emph {et~al.}(2023)\citenamefont {Xu}, \citenamefont {Xing}, \citenamefont {Xue}, \citenamefont {Lu}, \citenamefont {Fan}, \citenamefont {Fan}, \citenamefont {Shum},\ and\ \citenamefont {Cong}}]{xu2023recent}%
  \BibitemOpen
  \bibfield  {author} {\bibinfo {author} {\bibfnamefont {G.}~\bibnamefont {Xu}}, \bibinfo {author} {\bibfnamefont {H.}~\bibnamefont {Xing}}, \bibinfo {author} {\bibfnamefont {Z.}~\bibnamefont {Xue}}, \bibinfo {author} {\bibfnamefont {D.}~\bibnamefont {Lu}}, \bibinfo {author} {\bibfnamefont {J.}~\bibnamefont {Fan}}, \bibinfo {author} {\bibfnamefont {J.}~\bibnamefont {Fan}}, \bibinfo {author} {\bibfnamefont {P.~P.}\ \bibnamefont {Shum}},\ and\ \bibinfo {author} {\bibfnamefont {L.}~\bibnamefont {Cong}},\ }\bibfield  {title} {\bibinfo {title} {Recent advances and perspective of photonic bound states in the continuum},\ }\href@noop {} {\bibfield  {journal} {\bibinfo  {journal} {Ultrafast Science}\ }\textbf {\bibinfo {volume} {3}},\ \bibinfo {pages} {0033} (\bibinfo {year} {2023})}\BibitemShut {NoStop}%
\end{thebibliography}%

\end{document}